\documentstyle[prd,floats,aps,epsf,twocolumn]{revtex}
\input epsf

\newcommand{\dofig}[4]
{\begin{figure}[htb]
\begin{center}
\leavevmode
\hbox{%
\epsfxsize=#2
\epsfbox{#1}}
\caption{#3}
\label{#4}
\end{center}
\end{figure}}

\def \met {{\,/\!\!\!\!E_{T}}}

\def \GeV {{\rm GeV}}

\def \begineq {\begin{equation}}
\def \endeq {\end{equation}}
\def \scriptP {\mbox{${\cal P}$}}
\def \gothicP {\tilde{\scriptP}}
\def \ltapprox {\,\raisebox{-0.6ex}{$\stackrel{<}{\sim}$}\,}

\def \Sherlock {{Sleuth}}
\def \hse {{\small{hse}}}
\newcommand {\abs}[1]{\mid \! #1 \! \mid}

\newcommand{\ttbar}{$t{\bar t}$}

 \newcommand{\EMMFA}{18.4$\pm$1.4}
 \newcommand{\EMMTT}{0.011$\pm$0.003}
 \newcommand{\EMMDY}{0.5$\pm$0.2}
 \newcommand{\EMMWW}{3.9$\pm$1.0}
 \newcommand{\EMMZT}{25.6$\pm$6.5}
 \newcommand{\EMMTOT}{48.5$\pm$7.6}
 
 \newcommand{\EMMJFA}{8.7$\pm$1.0}
 \newcommand{\EMMJTT}{0.4$\pm$0.1}
 \newcommand{\EMMJDY}{0.1$\pm$0.03}
 \newcommand{\EMMJWW}{1.1$\pm$0.3}
 \newcommand{\EMMJZT}{3$\pm$0.8}
 \newcommand{\EMMJTOT}{13.2$\pm$1.5}
 
 \newcommand{\EMMJJFA}{2.7$\pm$0.6}
 \newcommand{\EMMJJTT}{1.8$\pm$0.5}
 \newcommand{\EMMJJDY}{0.012$\pm$0.006}
 \newcommand{\EMMJJWW}{0.18$\pm$0.05}
 \newcommand{\EMMJJZT}{0.5$\pm$0.2}
 \newcommand{\EMMJJTOT}{5.2$\pm$0.8}
 
 \newcommand{\EMMJJJFA}{0.4$\pm$0.2}
 \newcommand{\EMMJJJTT}{0.7$\pm$0.2}
 \newcommand{\EMMJJJDY}{0.005$\pm$0.004}
 \newcommand{\EMMJJJWW}{0.032$\pm$0.009}
 \newcommand{\EMMJJJZT}{0.07$\pm$0.05}
 \newcommand{\EMMJJJTOT}{1.3$\pm$0.3}

 \newcommand{\EMXFA}{30.2$\pm$1.8}
 \newcommand{\EMXTT}{3.1$\pm$0.5}
 \newcommand{\EMXDY}{0.7$\pm$0.1}
 \newcommand{\EMXWW}{5.2$\pm$0.8}
 \newcommand{\EMXZT}{29.2$\pm$4.5}
 \newcommand{\EMXTOT}{68.3$\pm$5.7}

\begin{document}

\onecolumn
\title{Search for New Physics in $e\mu X$ Data at D\O\ Using \Sherlock: A Quasi-Model-Independent Search Strategy for New Physics}
%
\author{                                                                      
B.~Abbott,$^{49}$                                                             
M.~Abolins,$^{46}$                                                            
V.~Abramov,$^{22}$                                                            
B.S.~Acharya,$^{15}$                                                          
D.L.~Adams,$^{56}$                                                            
M.~Adams,$^{33}$                                                              
G.A.~Alves,$^{2}$                                                             
N.~Amos,$^{45}$                                                               
E.W.~Anderson,$^{38}$                                                         
M.M.~Baarmand,$^{51}$
% ...
\\                                                                            
\vskip 0.30cm                                                                 
\centerline{(D\O\ Collaboration)}                                             
\vskip 0.30cm                                                                 
}                                                                             
\address{                                                                     
\centerline{$^{1}$Universidad de Buenos Aires, Buenos Aires, Argentina}       
\centerline{$^{2}$LAFEX, Centro Brasileiro de Pesquisas F{\'\i}sicas,         
                  Rio de Janeiro, Brazil}                                     
\centerline{$^{3}$Universidade do Estado do Rio de Janeiro,                   
                  Rio de Janeiro, Brazil}                                     
\centerline{$^{4}$Institute of High Energy Physics, Beijing,                  
                  People's Republic of China}                                 
\centerline{$^{5}$Universidad de los Andes, Bogot\'{a}, Colombia}             
\centerline{$^{6}$Charles University, Prague, Czech Republic}                 
\centerline{$^{7}$Institute of Physics, Academy of Sciences, Prague,          
                  Czech Republic}                                             
\centerline{$^{8}$Universidad San Francisco de Quito, Quito, Ecuador}         
\centerline{$^{9}$Institut des Sciences Nucl\'eaires, IN2P3-CNRS,             
                  Universite de Grenoble 1, Grenoble, France}                 
\centerline{$^{10}$CPPM, IN2P3-CNRS, Universit\'e de la M\'editerran\'ee,     
                  Marseille, France}                                          
\centerline{$^{11}$LPNHE, Universit\'es Paris VI and VII, IN2P3-CNRS,         
                  Paris, France}                                              
\centerline{$^{12}$DAPNIA/Service de Physique des Particules, CEA, Saclay,    
                  France}                                                     
\centerline{$^{13}$Panjab University, Chandigarh, India}                      
\centerline{$^{14}$Delhi University, Delhi, India}                            
\centerline{$^{15}$Tata Institute of Fundamental Research, Mumbai, India}     
\centerline{$^{16}$Seoul National University, Seoul, Korea}                   
\centerline{$^{17}$CINVESTAV, Mexico City, Mexico}                            
\centerline{$^{18}$University of Nijmegen/NIKHEF, Nijmegen, The Netherlands}                                                                                        
\centerline{$^{19}$Institute of Nuclear Physics, Krak\'ow, Poland}            
\centerline{$^{20}$Institute for Theoretical and Experimental Physics,        
                   Moscow, Russia}                                            
\centerline{$^{21}$Moscow State University, Moscow, Russia}                   
\centerline{$^{22}$Institute for High Energy Physics, Protvino, Russia}       
\centerline{$^{23}$Lancaster University, Lancaster, United Kingdom}           
\centerline{$^{24}$University of Arizona, Tucson, Arizona 85721}              
\centerline{$^{25}$Lawrence Berkeley National Laboratory and University of    
                  California, Berkeley, California 94720}                     
\centerline{$^{26}$University of California, Davis, California 95616}         
\centerline{$^{27}$California State University, Fresno, California 93740}     
\centerline{$^{28}$University of California, Irvine, California 92697}        
\centerline{$^{29}$University of California, Riverside, California 92521}     
\centerline{$^{30}$Florida State University, Tallahassee, Florida 32306}      
\centerline{$^{31}$University of Hawaii, Honolulu, Hawaii 96822}              
\centerline{$^{32}$Fermi National Accelerator Laboratory, Batavia,            
                   Illinois 60510}                                            
\centerline{$^{33}$University of Illinois at Chicago, Chicago,                
                   Illinois 60607}                                            
\centerline{$^{34}$Northern Illinois University, DeKalb, Illinois 60115}      
\centerline{$^{35}$Northwestern University, Evanston, Illinois 60208}         
\centerline{$^{36}$Indiana University, Bloomington, Indiana 47405}            
\centerline{$^{37}$University of Notre Dame, Notre Dame, Indiana 46556}       
\centerline{$^{38}$Iowa State University, Ames, Iowa 50011}                   
\centerline{$^{39}$University of Kansas, Lawrence, Kansas 66045}              
\centerline{$^{40}$Kansas State University, Manhattan, Kansas 66506}          
\centerline{$^{41}$Louisiana Tech University, Ruston, Louisiana 71272}        
\centerline{$^{42}$University of Maryland, College Park, Maryland 20742}      
\centerline{$^{43}$Boston University, Boston, Massachusetts 02215}            
\centerline{$^{44}$Northeastern University, Boston, Massachusetts 02115}      
\centerline{$^{45}$University of Michigan, Ann Arbor, Michigan 48109}         
\centerline{$^{46}$Michigan State University, East Lansing, Michigan 48824}   
\centerline{$^{47}$University of Nebraska, Lincoln, Nebraska 68588}           
\centerline{$^{48}$Columbia University, New York, New York 10027}             
\centerline{$^{49}$New York University, New York, New York 10003}             
\centerline{$^{50}$University of Rochester, Rochester, New York 14627}        
\centerline{$^{51}$State University of New York, Stony Brook,                 
                   New York 11794}                                            
\centerline{$^{52}$Brookhaven National Laboratory, Upton, New York 11973}     
\centerline{$^{53}$Langston University, Langston, Oklahoma 73050}             
\centerline{$^{54}$University of Oklahoma, Norman, Oklahoma 73019}            
\centerline{$^{55}$Brown University, Providence, Rhode Island 02912}          
\centerline{$^{56}$University of Texas, Arlington, Texas 76019}               
\centerline{$^{57}$Texas A\&M University, College Station, Texas 77843}       
\centerline{$^{58}$Rice University, Houston, Texas 77005}                     
\centerline{$^{59}$University of Washington, Seattle, Washington 98195}       
}                                                                             

\maketitle
\vskip 10pt

{\samepage
{\bf
\begin{center}
Abstract
\end{center}
}
\begin{center}
\begin{minipage}{.8\textwidth}
{\small We present a quasi-model-independent search for the physics responsible for electroweak symmetry breaking.  We define final states to be studied, and construct a rule that identifies a set of relevant variables for any particular final state.  A new algorithm (``\Sherlock'') searches for regions of excess in those variables and quantifies the significance of any detected excess.  After demonstrating the sensitivity of the method, we apply it to the semi-inclusive channel $e\mu X$ collected in 108 pb$^{-1}$ of $p\bar{p}$ collisions at $\sqrt{s} = 1.8$~TeV at the D\O\ experiment during 1992--1996 at the Fermilab Tevatron.  We find no evidence of new high $p_T$ physics in this sample.}
\end{minipage}
\end{center}
}

\vskip 6.0in
\begin{center}
\end{center}

\twocolumn

\clearpage
\tableofcontents

\section{Introduction}
\label{section:introduction}

It is generally recognized that the standard model, an extremely successful description of the fundamental particles and their interactions, must be incomplete.  Although there is likely to be new physics beyond the current picture, the possibilities are sufficiently broad that the first hint could appear in any of many different guises.  This suggests the importance of performing searches that are as model-independent as possible.

The word ``model'' can connote varying degrees of generality.  It can mean a particular model together with definite choices of parameters [e.g., mSUGRA~\cite{mSUGRA} with specified $m_{1/2}$, $m_{0}$, $A_0$, $\tan{\beta}$, and sign($\mu$)]; it can mean a particular model with unspecified parameters (e.g., mSUGRA); it can mean a more general model (e.g., SUGRA); it can mean an even more general model (e.g., gravity-mediated supersymmetry); it can mean a class of general models (e.g., supersymmetry); or it can be a set of classes of general models (e.g., theories of electroweak symmetry breaking).  As one ascends this hierarchy of generality, predictions of the ``model'' become less precise.  While there have been many searches for phenomena predicted by models in the narrow sense, there have been relatively few searches for predictions of the more general kind.

In this article we describe an explicit prescription 
for searching for the physics responsible for stabilizing electroweak symmetry breaking, in a manner that relies only upon what we are sure we know about electroweak symmetry breaking:  that its natural scale is on the order of the Higgs mass~\cite{HiggsVacuumExpectationValue}.  When we wish to emphasize the generality of the approach, we say that it is quasi-model-independent, where the ``quasi'' refers to the fact that the correct model of electroweak symmetry breaking should become manifest at the scale of several hundred GeV.

New sources of physics will in general lead to an excess over the expected background in some final state.  A general signature for new physics is therefore a region of variable space in which the probability for the background to fluctuate up to or above the number of observed events is small.  Because the mass scale of electroweak symmetry breaking is larger than the mass scale of most standard model backgrounds, we expect this excess to populate regions of high transverse momentum ($p_T$).  The method we will describe involves a systematic search for such excesses (although with a small modification it is equally applicable to searches for deficits).  Although motivated by the problem of electroweak symmetry breaking, this method is generally sensitive to any new high $p_T$ physics.

An important benefit of a precise {\it a priori} algorithm of the type we construct is that it allows an {\it a posteriori} evaluation of the significance of a small excess, in addition to providing a recipe for searching for such an effect.  The potential benefit of this feature can be seen by considering the two curious events seen by the CDF collaboration in their semi-inclusive $e\mu$ sample~\cite{CDFemuEvents} and one event in the data sample we analyze in this article, which have prompted efforts to determine the probability that the standard model alone could produce such a result~\cite{BarnettAndHall}.  This is quite difficult to do {\it a posteriori}, as one is forced to somewhat arbitrarily decide what is meant by ``such a result.''  The method we describe provides an unbiased and quantitative answer to such questions.  

``\Sherlock,'' a quasi-model-independent prescription for searching for high $p_T$ physics beyond the standard model, has two components:  
\begin{itemize}
\item{the definitions of physical objects and final states, and the variables relevant for each final state; and}
\item{an algorithm that systematically hunts for an excess in the space of those variables, and quantifies the likelihood of any excess found.}
\end{itemize}
We describe the prescription in Secs.~\ref{section:SearchStrategy} and~\ref{section:theAlgorithm}.  In Sec.~\ref{section:SearchStrategy} we define the physical objects and final states, and we construct a rule for choosing variables relevant for any final state.  In Sec.~\ref{section:theAlgorithm} we describe an algorithm that searches for a region of excess in a multidimensional space, and determines how unlikely it is that this excess arose simply from a statistical fluctuation, taking account of the fact that the search encompasses many regions of this space.  This algorithm is especially useful when applied to a large number of final states.  For a first application of \Sherlock, we choose the semi-inclusive $e\mu$ data set ($e\mu X$) because it contains ``known'' signals (pair production of $W$ bosons and top quarks) that can be used to quantify the sensitivity of the algorithm to new physics, and because this final state is prominent in several models of physics beyond the standard model~\cite{SUSYCharginoNeutralino,PatiSalamLeptoquarks}.  In Sec.~\ref{section:theDataSet} we describe the data set and the expected backgrounds from the standard model and instrumental effects.  In Sec.~\ref{section:sensitivity} we demonstrate the sensitivity of the method by ignoring the existence of top quark and $W$ boson pair production, and showing that the method can find these signals in the data.  In Sec.~\ref{section:Results} we apply the \Sherlock\ algorithm to the $e\mu X$ data set assuming the known backgrounds, including $WW$ and $t\bar{t}$, and present the results of a search for new physics beyond the standard model.

\section{Search strategy}
\label{section:SearchStrategy}

Most recent searches for new physics have followed a well-defined set of steps:  first selecting a model to be tested against the standard model, then finding a measurable prediction of this model that differs as much as possible from the prediction of the standard model, and finally comparing the predictions to \mbox{data}.  This is clearly the procedure to follow for a small number of compelling candidate theories.  Unfortunately, the resources required to implement this procedure grow almost linearly with the number of theories.  Although broadly speaking there are currently only three models with internally consistent methods of electroweak symmetry breaking --- supersymmetry~\cite{SusyReview}, strong dynamics~\cite{StrongDynamicsReview}, and theories incorporating large extra dimensions~\cite{LargeExtraDimensionsReview} --- the number of specific models (and corresponding experimental signatures) is in the hundreds.  Of these many specific models, at most one is a correct description of nature.

Another issue is that the results of searches for new physics can be unintentionally biased because the number of events under consideration is small, and the details of the analysis are often not specified before the data are examined.  An {\em a priori} technique would permit a detailed study without fear of biasing the result.

We first specify the prescription in a form that should be applicable to any collider experiment sensitive to physics at the electroweak scale.  We then provide aspects of the prescription that are specific to D\O.  Other experiments wishing to use this prescription would specify similar details appropriate to their detectors.

\subsection{General prescription}

We begin by defining final states, and follow by motivating the variables we choose to consider for each of those final states.  We assume that standard particle identification requirements, often detector-specific, have been agreed upon.  The understanding of all backgrounds, through Monte Carlo programs and data, is crucial to this analysis, and requires great attention to detail.  Stan\-dard methods for understanding backgrounds --- comparing different Monte Carlos, normalizing background predictions to observation, obtaining instrumental backgrounds from related samples, demonstrating agreement in limited regions of variable space, and calibrating against known physical quantities, among many others --- are needed and used in this analysis as in any other.  Uncertainties in backgrounds, which can limit the sensitivity of the search, are naturally folded into this approach.  

\subsubsection{Final states}

In this subsection we partition the data into final states.  The specification is based on the notions of exclusive channels and standard particle identification.

\paragraph{Exclusiveness.}

Although analyses are frequently performed on inclusive samples, considering only exclusive final states has several advantages in the context of this approach:
\begin{itemize}
\item{the presence of an extra object (electron, photon, muon, \ldots) in an event often qualitatively affects the probable interpretation of the event;}
\item{the presence of an extra object often changes the variables that are chosen to characterize the final state; and}
\item{using inclusive final states can lead to ambiguities when different channels are combined.}
\end{itemize}
We choose to partition the data into exclusive categories.

\paragraph{Particle identification.}

We now specify the labeling of these exclusive final states.  The general principle is that we label the event as completely as possible, as long as we have a high degree of confidence in the label.  This leads naturally to an explicit prescription for labeling final states.

Most multipurpose experiments are able to identify electrons, muons, photons, and jets, and so we begin by considering a final state to be described by the number of isolated electrons, muons, photons, and jets observed in the event, and whether there is a significant imbalance in transverse momentum ($\met$).  We treat $\met$ as an object in its own right, which must pass certain quality \mbox{criteria}.  If $b$-tagging, $c$-tagging, or $\tau$-tagging is possible, then we can differentiate among jets arising from $b$ quarks, $c$ \mbox{quarks}, light quarks, and hadronic tau decays.  If a magnetic field can be used to obtain the electric charge of a lepton,  we split the charged leptons $\ell$ into $\ell^+$ and $\ell^-$ but consider final states that are related through global charge conjugation to be equivalent in $p\bar{p}$ or $e^+e^-$ (but not $pp$) collisions.  Thus $e^+e^-\gamma$ is a different final state than $e^+e^+\gamma$, but $e^+e^+\gamma$ and $e^-e^-\gamma$ together make up a single final state.  The definitions of these objects are logically specified for general use in all analyses, and we use these standard identification criteria to define our objects.

We can further specify a final state by identifying any $W$ or $Z$ bosons in the event.  This has the effect (for example) of splitting the $eejj$, $\mu\mu jj$, and $\tau\tau jj$ final states into the $Zjj$, $eejj$, $\mu\mu jj$, and $\tau\tau jj$ channels, and splitting the $e \met j j$, $\mu \met j j$, and $\tau \met j j$ final states into  $W j j$, $e \met j j$, $\mu \met j j$, and $\tau \met j j$ channels. 

We combine a $\ell^+\ell^-$ pair into a $Z$ if their invariant mass $M_{\ell^+\ell^-}$ falls within a $Z$ boson mass window ($82 \leq M_{\ell^+\ell^-} \leq 100$~GeV for D\O\ data) and the event contains neither significant $\met$ nor a third charged lepton.  If the event contains exactly one photon in addition to a $\ell^+\ell^-$ pair, and contains neither significant $\met$ nor a third charged lepton, and if $M_{\ell^+\ell^-}$ does not fall within the $Z$ boson mass window, but $M_{\ell^+\ell^- \gamma}$ does, then the $\ell^+\ell^-\gamma$ triplet becomes a $Z$ boson.  If the experiment is not capable of distinguishing between $\ell^+$ and $\ell^-$ and the event contains exactly two $\ell$'s, they are assumed to have opposite charge.  A lepton and $\met$ become a $W$ boson if the transverse mass $M^T_{\ell\met}$ is within a $W$ boson mass window ($30 \leq M^T_{\ell\met} \leq 110$~GeV for D\O\ data) and the event contains no second charged lepton.  Because the $W$ boson mass window is so much wider than the $Z$ boson mass window, we make no attempt to identify radiative $W$ boson decays.

We do not identify top quarks, gluons, nor $W$ or $Z$ bosons from hadronic decays because we would have little confidence in such a label.  Since the predicted cross sections for new physics are comparable to those for the production of detectable $ZZ$, $WZ$, and $WW$ final states, we also elect not to identify these final states.  

\paragraph{Choice of final states to study.}

Because it is not realistic to specify backgrounds for all possible exclusive final states, choosing prospective final states is an important issue.  Theories of physics beyond the standard model make such wide-ranging predictions that neglect of any particular final state purely on theoretical grounds would seem unwise.  Focusing on final states in which the data themselves suggest something interesting can be done without fear of bias if all final states and variables for those final states are defined prior to examining the data.  Choosing variables is the subject of the next section.

\subsubsection{Variables}

We construct a mapping from each final state to a list of key variables for that final state using a simple, well-motivated, and short set of rules.  The rules, which are summarized in Table~\ref{tbl:VariableRules}, are obtained through the following reasoning:

\begin{itemize}

\item{There is strong reason to believe that the physics responsible for electroweak symmetry breaking occurs at the scale of the mass of the Higgs boson, or on the order of a few hundred GeV.  Any new massive particles associated with this physics can therefore be expected to decay into objects with large transverse momenta in the final state.}

\item{Many models of electroweak symmetry breaking predict final states with large missing transverse energy.  This arises in a large class of $R$-parity conserving supersymmetric theories containing a neutral, stable, lightest supersymmetric particle; in theories with ``large'' extra dimensions containing a Kaluza-Klein tower of gravitons that escape into the multidimensional ``bulk space''~\cite{LargeExtraDimensionsReview}; and more generally from neutrinos produced in electroweak boson decay.  If the final state con\-tains significant $\met$, then $\met$ is included in the list of promising variables.  We do not use $\met$ that is reconstructed as a $W$ boson decay product, following the prescription for $W$ and $Z$ boson identification outlined above.}

\item{If the final state contains one or more leptons we use the summed scalar transverse momenta $\sum{p_T^\ell}$, where the sum is over all leptons whose identity can be determined and whose momenta can be accurately measured.  Leptons that are reconstructed as $W$ or $Z$ boson decay products are not included in this sum, again following the prescription for $W$ and $Z$ boson identification outlined above.  We combine the momenta of $e$, $\mu$, and $\tau$ leptons because these objects are expected to have comparable transverse momenta on the basis of lepton universality in the standard model and the negligible values of lepton masses.}

\item{Similarly, photons and $W$ and $Z$ bosons are most likely to signal the presence of new phenomena when they are produced at high transverse momentum.  Since the expected transverse momenta of the electroweak gauge bosons are comparable, we use the variable $\sum{p_T^{\gamma/W/Z}}$, where the scalar sum is over all electroweak gauge bosons in the event, for final states with one or more of them identified.}

\item{For events with one jet in the final state, the transverse energy of that jet is an important variable.  For events with two or more jets in the final state, previous analyses have made use of the sum of the transverse energies of all but the leading jet~\cite{topQuarkMass}.  The reason for excluding the energy of the leading jet from this sum is that while a hard jet is often obtained from QCD radiation, hard second and third radiative jets are relatively much less likely.  We therefore choose the variable $\sum'{p_T^j}$ to describe the jets in the final state, where $\sum'{p_T^j}$ denotes $p_T^{j_1}$ if the final state contains only one jet, and $\sum_{i=2}^n{p_T^{j_i}}$ if the final state contains two or more jets.  Since QCD dijets are a large background in all-jets final states, $\sum'{p_T^j}$ refers instead to $\sum_{i=3}^n{p_T^{j_i}}$ for final states containing $n$ jets and nothing else, where $n \ge 3$.}

\end{itemize}

When there are exactly two objects in an event (e.g., one $Z$ boson and one jet), their $p_T$ values are expected to be nearly equal, and we therefore use the average $p_T$ of the two objects.  When there is only one object in an event (e.g., a single $W$ boson), we use no variables, and simply perform a counting experiment.  

Other variables that can help pick out specific signatures can also be defined.  Although variables such as invariant mass, angular separation between particular final state objects, and variables that characterize event topologies may be useful in testing a particular model, these variables tend to be less powerful in a general search.  Appendix~\ref{section:MoreOnVariables} contains a more detailed discussion of this point.  In the interest of keeping the list of variables as general, well-motivated, powerful, and short as possible, we elect to stop with those given in Table~\ref{tbl:VariableRules}.  We expect evidence for new physics to appear in the high tails of the $\met$, $\sum{p_T^\ell}$, $\sum{p_T^{\gamma/W/Z}}$, and $\sum'{p_T^j}$ distributions.

\begin{table}[htb]
\centering
\begin{tabular}{cc}
If the final state includes & then consider the variable \\ \hline
$\met$ & $\met$ \\ 
one or more charged leptons & $\sum{p_T^\ell}$ \\ 
one or more electroweak bosons & $\sum{p_T^{\gamma/W/Z}}$ \\ 
one or more jets & $\sum'{p_T^j}$ \\ 
\end{tabular}
\caption{A quasi-model-independently motivated list of interesting variables for any final state.  The set of variables to consider for any particular final state is the union of the variables in the second column for each row that pertains to that final state.  Here $\ell$ denotes $e$, $\mu$, or $\tau$.  The notation $\sum'{p_T^j}$ is shorthand for $p_T^{j_1}$ if the final state contains only one jet, $\sum_{i=2}^n{p_T^{j_i}}$ if the final state contains $n \geq 2$ jets, and $\sum_{i=3}^n{p_T^{j_i}}$ if the final state contains $n$ jets and nothing else, with $n \geq 3$.  Leptons and missing transverse energy that are reconstructed as decay products of $W$ or $Z$ bosons are not considered separately in the left-hand column.}
\label{tbl:VariableRules}
\end{table}

\subsection{Search strategy:  D\O\ Run I}
\label{section:D0RunISearchStrategySupplement}

The general search strategy just outlined is applicable to any collider experiment searching for the physics responsible for electroweak symmetry breaking.  Any particular experiment that wishes to use this strategy needs to specify object and variable definitions that reflect the capabilities of the detector.  This section serves this function for the D\O\ detector~\cite{D0Detector} in its 1992--1996 run (Run I) at the Fermilab Tevatron.  Details in this subsection supersede those in the more general section above.

\subsubsection{Object definitions}

The particle identification algorithms used here for electrons, muons, jets, and photons are similar to those used in many published D\O\ analyses. We summarize them here.

\paragraph{Electrons.}

D\O\ had no central magnetic field in Run I; therefore, there is no way to distinguish between electrons and positrons.  Electron candidates with transverse energy greater than 15 GeV, within the fiducial region of $\abs{\eta}<1.1$ or $1.5<\abs{\eta}<2.5$ (where $\eta=-\ln\tan(\theta/2)$, with $\theta$ the polar angle with respect to the colliding proton's direction), and satisfying standard electron identification and isolation requirements as defined in Ref.~\cite{topCrossSection} are accepted.

\paragraph{Muons.}

We do not distinguish between positively and negatively charged muons in this analysis.  We accept muons with transverse momentum greater than 15 GeV and $\abs{\eta}<1.7$ that satisfy standard muon identification and isolation requirements~\cite{topCrossSection}.

\paragraph{$\met$.}
\label{section:D0RunImet}

The missing transverse energy, $\met$, is the energy required to balance the measured energy in the event.  In the calorimeter, we calculate
\begineq
\met^{\!\!\!\!\rm cal} = \abs{\sum_i{E_i \sin{\theta_i} ( \cos{\phi_i}\,\hat{x} + \sin{\phi_i}\,\hat{y})}},
\endeq
where $i$ runs over all calorimeter cells, $E_i$ is the energy deposited in the $i^{th}$ cell, and $\phi_i$ is the azimuthal and $\theta_i$ the polar angle of the center of the $i^{th}$ cell, measured with respect to the event vertex.

An event is defined to contain a $\met$ ``object'' only if we are confident that there is significant missing transverse energy.  Events that do not contain muons are said to contain $\met$ if $\met^{\!\!\!\!\rm cal} > 15~\GeV$.  Using track deflection in magnetized steel toroids, the muon momentum resolution in Run I is
\begineq
\delta(1/p) = 0.18 (p-2)/p^2 \oplus 0.003,
\endeq
 where $p$ is in units of GeV, and the $\oplus$ means addition in quadrature.  This is significantly coarser than the electromagnetic and jet energy resolutions, parameterized by
\begineq
{\delta E}/E = 15\%/\sqrt{E} \oplus 0.3\%
\endeq
and
\begineq
{\delta E}/E = 80\%/\sqrt{E},
\endeq
respectively.  Events that contain exactly one muon are deemed to contain $\met$ on the basis of muon number conservation rather than on the basis of the muon momentum measurement.  We do not identify a $\met$ object in events that contain two or more muons.

\paragraph{Jets.}

Jets are reconstructed in the calorimeter using a fixed-size cone algorithm, with a cone size of $\Delta R=\sqrt{(\Delta\phi)^2+(\Delta\eta)^2} = 0.5$~\cite{searchForTopPRD}. We require jets to have $E_T > 15~\GeV$ and $\abs{\eta} < 2.5$.  We make no attempt to distinguish among light quarks, gluons, charm quarks, bottom quarks, and hadronic tau decays.
 
\paragraph{Photons.}

Isolated photons that pass standard \mbox{identification} requirements~\cite{diphotonLargeMet}, have transverse energy greater than 15~GeV, and are in the fiducial region $\abs{\eta}<1.1$ or $1.5<\abs{\eta}<2.5$ are labeled photon objects.

\paragraph{$W$ bosons.}

Following the general prescription described above, an electron (as defined above) and $\met$ become a $W$ boson if their transverse mass is within the $W$ boson mass window ($30 \leq M^T_{\ell\met} \leq 110$~GeV), and the event contains no second charged lepton.  Because the muon momentum measurement is coarse, we do not use a transverse mass window for muons.  From Sec.~\ref{section:D0RunImet}, any event containing a single muon is said to also contain $\met$; thus any event containing a muon and no second charged lepton is said to contain a $W$ boson.

\paragraph{$Z$ bosons.}
\label{section:Z}

We use the rules in the previous section for combining an $ee$ pair or $ee\gamma$ triplet into a $Z$ boson.  We do not attempt to reconstruct a $Z$ boson in events containing three or more charged leptons.  For events containing two muons and no third charged lepton, we fit the event to the hypothesis that the two muons are decay products of a $Z$ boson and that there is no $\met$ in the event.  If the fit is acceptable, the two muons are considered to be a $Z$ boson.

\subsubsection{Variables}

The variables provided in the general prescription above also need minor revision to be appropriate for the D\O\ experiment.  

\paragraph{$\sum{p_T^\ell}$.}

We do not attempt to identify $\tau$ leptons, and the momentum resolution for muons is coarse.  For events that contain no leptons other than muons, we define $\sum{p_T^\ell} = \sum{p_T^\mu}$.  For events that contain one or more electrons, we define $\sum{p_T^\ell} = \sum{p_T^e}$. This is identical to the general definition provided above except for events containing both one or more electrons and one or more muons.  In this case, we have decided to define $\sum{p_T^\ell}$ as the sum of the momenta of the electrons only, rather than combining the well-measured electron momenta with the poorly-measured muon momenta.

\paragraph{$\met$.}
\label{section:Variables:met}

$\met$ is defined by $\met = \met^{\!\!\!\!\rm cal}$, where $\met^{\!\!\!\!\rm cal}$ is the missing transverse energy as summed in the calorimeter.  This sum includes the $p_T$ of electrons, but only a negligible fraction of the $p_T$ of muons.

\paragraph{$\sum{p_T^{\gamma/W/Z}}$.}

We use the definition of $\sum{p_T^{\gamma/W/Z}}$ provided in the general prescription: the sum is over all electroweak gauge bosons in the event, for final states with one or more of them.  We note that if a $W$ boson is formed from a $\mu$ and $\met$, then $p_T^W = \met^{\!\!\!\!\rm cal}$.

\section{Sleuth algorithm}
\label{section:theAlgorithm}

Given a data sample, its final state, and a set of variables appropriate to that final state, we now describe the algorithm that determines the most interesting region in those variables and quantifies the degree of interest.

\subsection{Overview}

Central to the algorithm is the notion of a ``region'' ($R$).  A region can be regarded simply as a volume in the variable space defined by Table~\ref{tbl:VariableRules}, satisfying certain special properties to be discussed in Sec.~\ref{section:Regions}.  The region contains $N$ data points and an expected number of background events $\hat{b}_R$.  We can consequently compute the weighted probability $p_N^R$, defined in Sec.~\ref{section:probabilities}, that the background in the region fluctuates up to or beyond the observed number of events.  If this probability is small, we flag the region as potentially interesting.

In any reasonably-sized data set, there will always be regions in which the probability for $b_R$ to fluctuate up to or above the observed number of events is small.  The relevant issue is how often this can happen in an ensemble of hypothetical similar experiments (\hse's).  This question can be answered by performing these hypothetical similar experiments; i.e., by generating random events drawn from the background distribution, finding the least probable region, and repeating this many times.  The fraction of hypothetical similar experiments that yields a probability as low as the one observed in the data provides the appropriate measure of the degree of interest.

Although the details of the algorithm are complex, the interface is straightforward.  What is needed is a data sample, a set of events for each background process $i$, and the number of background events $\hat{b}_i \pm \delta \hat{b}_i$ from each background process expected in the data sample.  The output gives the region of greatest excess and the fraction of hypothetical similar experiments that would yield such an excess.

The algorithm consists of seven steps:

\begin{enumerate}
\item{Define regions $R$ about any chosen set of $N=1,\ldots,N_{\rm data}$ data points in the sample of $N_{\rm data}$ data points.}
\item{Estimate the background $\hat{b}_R$ expected within these $R$.}
\item{Calculate the weighted probabilities $p_N^R$ that $b_R$ can fluctuate to $\ge N$.}
\item{For each $N$, determine the $R$ for which $p_N^R$ is minimum.  Define $p_N = \min_R{(p_N^R)}$.}
\item{Determine the fraction $P_N$ of hypothetical similar experiments in which the $p_N$(\hse) is smaller than the observed $p_N$(data).}
\item{Determine the $N$ for which $P_N$ is minimized.  Define $P=\min_N{(P_N)}$.}
\item{Determine the fraction $\cal P$ of hypothetical similar experiments in which the $P$(\hse) is smaller than the observed $P$(data).}
\end{enumerate}
Our notation is such that a lowercase $p$ represents a probability, while an uppercase $P$ or $\cal P$ represents the fraction of hypothetical similar experiments that would yield a less probable outcome.  The symbol representing the minimization of $p_N^R$ over $R$, $p_N$ over $N$, or $P_N$ over $N$ is written without the superscript or subscript representing the varied property (i.e., $p_N$, $p$, or $P$, respectively).  The rest of this section discusses these steps in greater detail.

\subsection{Steps 1 and 2:  Regions}
\label{section:Regions}

When there are events that do not appear to follow some expected distribution, such as the event at $x=61$ in Fig.~\ref{fig:pointOnTail}, we often attempt to estimate the probability that the event is consistent with coming from that distribution.  This is generally done by choosing some region around the event (or an accumulation of events), integrating the background within that region, and computing the probability that the expected number of events in that region could have fluctuated up to or beyond the observed number.

Of course, the calculated probability depends on how the region containing the events is chosen.  If the region about the event is infinitesimal, then the expected number of background events in the region (and therefore this probability) can be made arbitrarily small.  A possible approach in one dimension is to define the region to be the interval bounded below by the point halfway between the interesting event and its nearest neighbor, and bounded above by infinity.  For the case shown in Fig.~\ref{fig:pointOnTail}, this region would be roughly the interval $\left(46,\infty\right)$.

{\dofig {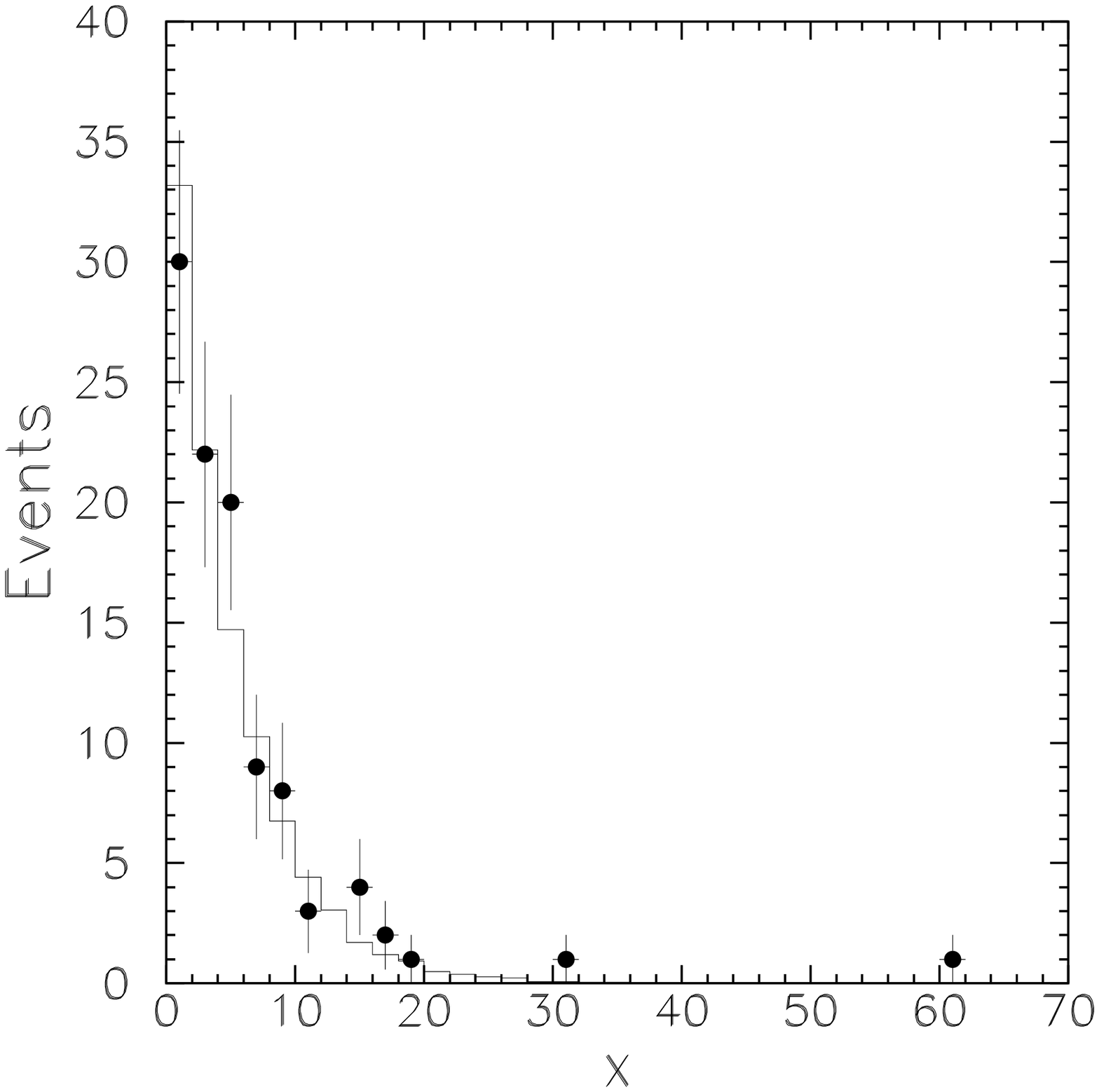} {3.0in} {Example of a data set with a potentially anomalous point.  The solid histogram is the expected distribution, and the points with error bars are the data.  The bulk of the data is well described by the background prediction, but the point located at $x=61$ appears out of place.} {fig:pointOnTail}}

Such a prescription breaks down in two or more dimensions, and it is not entirely satisfactory even in one dimension.  In particular, it is not clear how to proceed if the excess occurs somewhere other than at the tail end of a distribution, or how to generalize the interval to a well-defined contour in several dimensions.  As we will see, there are significant advantages to having a precise definition of a region about a potentially interesting set of data points.  This is provided in Sec.~\ref{section:Voronoi}, after we specify the variable space itself.

\subsubsection{Variable transformation}
\label{section:VariableTransformation}

Unfortunately, the region that we choose about the point on the tail of Fig.~\ref{fig:pointOnTail} changes if the variable is some function of $x$, rather than $x$ itself.  If the region about each data point is to be the subspace that is closer to that point than to any other one in the sample, it would therefore be wise to minimize any dependence of the selection on the shape of the background distribution.  For a background distributed uniformly between 0 and 1 (or, in $d$ dimensions, uniform within the unit ``box'' $\left[0,1\right]^d$), it is reasonable to define the region associated with an event as the variable subspace closer to that event than to any other event in the sample.  If the background is not already uniform within the unit box, we transform the variables so that it becomes uniform.  The details of this transformation are provided in Appendix~\ref{section:VariableTransformationAppendix}.

With the background distribution trivialized, the rest of the analysis can be performed within the unit box without worrying about the background shape.  A considerable simplification is therefore achieved through this transformation.  The task of determining the expected background within each region, which would have required a Monte Carlo integration of the background distribution over the region, reduces to the problem of determining the volume of each region.  The problem is now completely specified by the transformed coordinates of the data points, the total number of expected background events $\hat{b}$, and its uncertainty $\delta \hat{b}$.

\subsubsection{Voronoi diagrams}
\label{section:Voronoi}

Having defined the variable space by requiring a uniform background distribution, we can now define more precisely what is meant by a region.  Figure~\ref{fig:VoronoiAntiCorner} shows a 2-dimensional variable space $V$ containing seven data points in a unit square.  For any $v \in V$, we say that $v$ {\it belongs to} the data point $D_i$ if $\abs{v-D_i}<\abs{v-D_j}$ for all $j\ne i$; that is, $v$ belongs to $D_i$ if $v$ is closer to $D_i$ than to any other data point.  In Fig.~\ref{fig:VoronoiAntiCorner}(a), for example, any $v$ lying within the variable subspace defined by the pentagon in the upper right-hand corner belongs to the data point located at $(0.9, 0.8)$. The set of points in $V$ that do not belong to any data point [those points on the lines in Fig.~\ref{fig:VoronoiAntiCorner}(a)] has zero measure and may be ignored.

We define a {\it region} around a set of data points in a variable space $V$ to be the set of all points in $V$ that are closer to one of the data points in that set than to any data points outside that set.  A region around a single data point is the union of all points in $V$ that belong to that data point, and is called a 1-region.  A region about a set of $N$ data points is the union of all points in $V$ that belong to any one of the data points, and is called an $N$-region; an example of a 2-region is shown as the shaded area in Fig.~\ref{fig:VoronoiAntiCorner}(b).  $N_{\rm{data}}$ data points thus partition $V$ into $N_{\rm{data}}$ 1-regions.  Two data points are said to be neighbors if their 1-regions share a border -- the points at $(0.75, 0.9)$ and $(0.9, 0.8)$ in Fig.~\ref{fig:VoronoiAntiCorner}, for example, are neighbors.  A diagram such as Fig.~\ref{fig:VoronoiAntiCorner}(a), showing a set of data points and their regions, is known as a {\it Voronoi diagram}.  We use a program called {\small HULL}~\cite{Hull} for this computation.

{\dofig {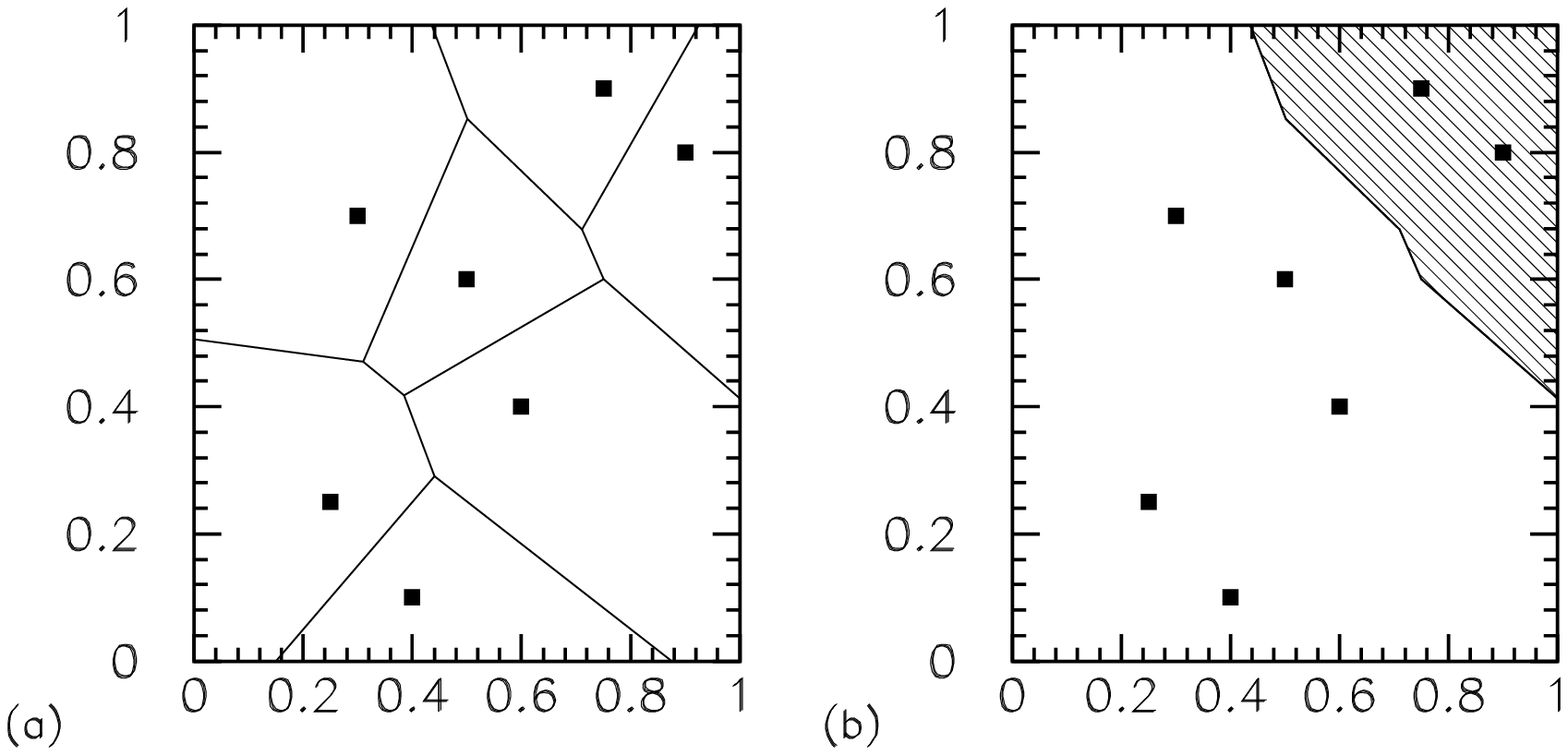} {3.0in} {A Voronoi diagram.  (a) The seven data points are shown as black dots; the lines partition the space into seven regions, with one region belonging to each data point.  (b) An example of a 2-region.} {fig:VoronoiAntiCorner}}

\subsubsection{Region criteria}
\label{section:RegionCriteria}

The explicit definition of a region that we have just provided reduces the number of contours we can draw in the variable space from infinite to a mere $2^{N_{\rm data}}-1$, since any region either contains all of the points belonging to the $i^{th}$ data event or it contains none of them.  In \mbox{fact}, because many of these regions have a shape that makes them implausible as ``discovery regions'' in which new physics might be concentrated, the number of possible regions may be reduced further.  For example, the region in Fig.~\ref{fig:VoronoiAntiCorner} containing only the lower-leftmost and the upper-rightmost data points is unlikely to be a discovery region, whereas the region shown in Fig.~\ref{fig:VoronoiAntiCorner}(b) containing the two upper-rightmost data points is more likely (depending upon the nature of the variables).

We can now impose whatever criteria we wish upon the regions that we allow \Sherlock\ to consider.  In general we will want to impose several criteria, and in this case we write the net criterion $c_R = c_R^1 c_R^2 \ldots$ as a product of the individual criteria, where $c_R^i$ is to be read ``the extent to which the region $R$ satisfies the criterion $c^i$.''  The quantities $c_R^i$ take on values in the interval $[0,1]$, where $c_R^i \rightarrow 0$ if $R$ badly fails $c^i$, and $c_R^i \rightarrow 1$ if $R$ easily satisfies $c^i$.

Consider as an example $c=$ {\em AntiCornerSphere}, a simple criterion that we have elected to impose on the regions in the $e\mu X$ sample.  Loosely speaking, a region $R$ will satisfy this criterion ($c_R \rightarrow 1$) if all of the data points inside the region are farther from the origin than all of the data points outside the region.  This situation is shown, for example, in Fig.~\ref{fig:VoronoiAntiCorner}(b).  For every event $i$ in the data set, denote by $r_i$ the distance of the point in the unit box to the origin, let $r'$ be $r$ transformed so that the background is uniform in $r'$ over the interval $[0,1]$, and let $r'_i$ be the values $r_i$ so transformed.  Then define 
\begineq
c_R = \left\{ 
\begin{array}{ll}
0 & , \left( \frac{1}{2} + \frac{r'^{\rm in}_{\rm min} - r'^{\rm out}_{\rm max}}{\xi} \right) < 0 \nonumber \\
\left( \frac{1}{2} + \frac{r'^{\rm in}_{\rm min} - r'^{\rm out}_{\rm max}}{\xi} \right) & , 0 \le \left( \frac{1}{2} + \frac{r'^{\rm in}_{\rm min} - r'^{\rm out}_{\rm max}}{\xi} \right) \le 1 \nonumber \\
1 & , 1 < \left( \frac{1}{2} + \frac{r'^{\rm in}_{\rm min} - r'^{\rm out}_{\rm max}}{\xi} \right)
\end{array} \right.
\endeq
 where $r'^{\rm in}_{\rm min} = \min_{i\in R}{(r'_i)}$, $r'^{\rm out}_{\rm max} = \max_{i\not\in R}{(r'_i)}$, and $\xi=1/(4 N_{\rm data})$ is an average separation distance between data points in the variable $r'$.

Notice that in the limit of vanishing $\xi$, the criterion $c$ becomes a boolean operator, returning ``true'' when all of the data points inside the region are farther from the origin than all of the data points outside the region, and ``false'' otherwise.  In fact, many possible criteria have a scale $\xi$ and reduce to boolean operators when $\xi$ vanishes.  This scale has been introduced to ensure continuity of the final result under small changes in the background estimate.  In this spirit, the ``extent to which $R$ satisfies the criterion $c$'' has an alternative interpretation as the ``fraction of the time $R$ satisfies the criterion $c$,'' where the average is taken over an ensemble of slightly perturbed background estimates and $\xi$ is taken to vanish, so that ``satisfies'' makes sense.  We will use $c_R$ in the next section to define an initial measure of the degree to which $R$ is interesting.

We have considered several other criteria that could be imposed upon any potential discovery region to ensure that the region is ``reasonably shaped'' and ``in a believable location.''  We discuss a few of these criteria in Appendix~\ref{section:RegionCriteriaAppendix}.

\subsection{Step 3:  Probabilities and uncertainties}

Now that we have specified the notion of a region, we can define a quantitative measure of the ``degree of interest'' of a region.  

\subsubsection{Probabilities}
\label{section:probabilities}

Since we are looking for regions of excess, the appropriate measure of the degree of interest is a slight modification of the probability of background fluctuating up to or above the observed number of events.  For an $N$-region $R$ in which $\hat{b}_R$ background events are expected and $\hat{b}_R$ is precisely known, this probability is
\begineq
\label{eqn:eqn4}
\sum_{i=N}^\infty{\frac{e^{-\hat{b}_R} (\hat{b}_R)^i}{i!}}.
\endeq
We use this to define the weighted probability
\begineq
\label{eqn:averageProbability}
p_N^R = \left( \sum_{i=N}^\infty{\frac{e^{-\hat{b}_R} (\hat{b}_R)^i}{i!}} \right) c_R + (1-c_R),
\endeq
which one can also think of as an ``average probability,'' where the average is taken over the ensemble of slightly perturbed background estimates referred to above.  By construction, this quantity has all of the properties we need:  it reduces to the probability in \mbox{Eq.}~\ref{eqn:eqn4} in the limit that $R$ easily satisfies the region criteria, it saturates at unity in the limit that $R$ badly fails the region criteria, and it exhibits continuous behavior under small perturbations in the background estimate between these two extremes. 

\subsubsection{Systematic uncertainties}

The expected number of events from each background process has a systematic uncertainty that must be taken into account.  There may also be an uncertainty in the shape of a particular background distribution --- for example, the tail of a distribution may have a larger systematic uncertainty than the mode.  

The background distribution comprises one or more contributing background processes.  For each background process we know the number of expected events and the systematic uncertainty on this number, and we have a set of Monte Carlo points that tell us what that background process looks like in the variables of interest.  A typical situation is sketched in Fig.~\ref{fig:fig25}.

{\dofig {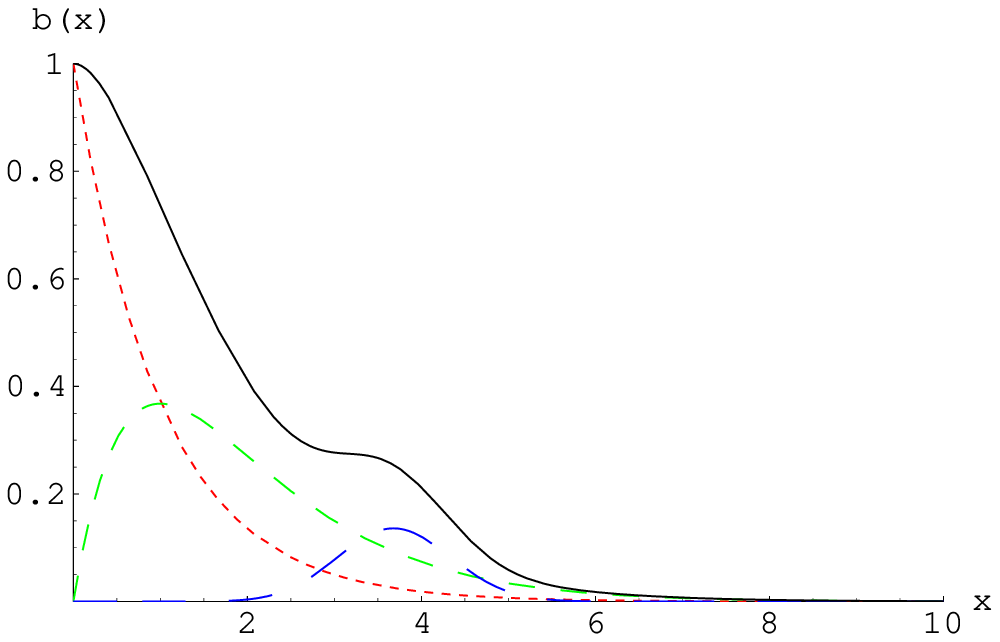} {3.0in} {An example of a one-dimensional background distribution with three sources.  The normalized shapes of the individual background processes are shown as the dashed lines; the solid line is their sum.  Typically, the normalizations for the background processes have separate systematic errors.  These errors can change the shape of the total background curve in addition to its overall normalization.  For example, if the long-dashed curve has a large systematic error, then the solid curve will be known less precisely in the region $(3,5)$ than in the region $(0,3)$ where the other two backgrounds dominate.} {fig:fig25}}

The multivariate transformation described in Sec.~\ref{section:VariableTransformation} is obtained assuming that the number of events expected from each background process is known precisely.  This fixes each event's position in the unit box, its neighbors, and the volume of the surrounding region.  The systematic uncertainty ${\delta\hat{b}_R}$ on the number of background events in a given region is computed by combining the systematic uncertainties for each individual background process.  Eq.~\ref{eqn:averageProbability} then generalizes to
\begin{eqnarray}
p_N^R = c_R \int_{0}^{\infty}{\sum_{i=N}^\infty{\frac{e^{-b} b^i}{i!} \frac{1}{\sqrt{2\pi}(\delta \hat{b}_R)}}} \times \nonumber \\ \exp\left(-\frac{(b - \hat{b}_R)^2}{2 (\delta \hat{b}_R)^2}\right) \, db \nonumber \\ + \, (1-c_R),
\end{eqnarray}
which is seen to reduce to Eq.~\ref{eqn:averageProbability} in the limit $\delta \hat{b}_R \rightarrow 0$.

This formulation provides a way to take account of systematic uncertainties on the shapes of distributions, as well.  For example, if there is a larger systematic uncertainty on the tail of a distribution, then the background process can be broken into two components, one describing the bulk of the distribution and one describing the tail, and a larger systematic uncertainty assigned to the piece that describes the tail.  Correlations among the various components may also be assigned.

We vary the number of events generated in the hypothetical similar experiments according to the systematic and statistical uncertainties.  The systematic errors are accounted for by pulling a vector of the ``true'' number of expected background events $\vec{b}$ from the distribution
\begineq
\label{eqn:dealingWithSystematicErrorsOnBackground}
p(\vec{b})= \frac{1}{\sqrt{2 \pi \abs{\Sigma}} } \exp\left({-\frac{1}{2}(b_i-\hat{b}_i)\Sigma^{-1}_{ij}(b_j-\hat{b}_j)}\right),
\endeq
where $\hat{b}_i$ is the number of expected background events from process $i$, as before, and $b_i$ is the $i^{th}$ component of $\vec{b}$.  We have introduced a covariance matrix $\Sigma$, which is diagonal with components $\Sigma_{ii}=(\delta\hat{b}_i)^2$ in the limit that the systematic uncertainties on the different background processes are uncorrelated, and we assume summation on repeated indices in Eq.~\ref{eqn:dealingWithSystematicErrorsOnBackground}.  The statistical uncertainties in turn are allowed for by choosing the number of events $N_i$ from each background process $i$ from the Poisson distribution
\begineq
\label{eqn:eqn7}
P(N_i) = \frac{e^{-b_i} b_i^{N_i}}{{N_i}!},
\endeq
where $b_i$ is the $i^{th}$ component of the vector $\vec{b}$ just determined.

\subsection{Step 4:  Exploration of regions}

Knowing how to calculate $p_N^R$ for a specific $N$-region $R$ allows us to determine which of two $N$-regions is more interesting.  Specifically, an $N$-region $R_1$ is more interesting than another $N$-region $R_2$ if $p_N^{R_1} < p_N^{R_2}$.  This allows us to compare regions of the same size (the same $N$), although, as we will see, it does not allow us to compare regions of different size.

Step 4 of the algorithm involves finding the most interesting $N$-region for each fixed $N$ between 1 and $N_{\rm data}$.  This most interesting $N$-region is the one that minimizes $p_N^R$, and these $p_N = \min_R(p_N^R)$ are needed for the next step in the algorithm.

Even for modestly sized problems (say, two di\-men\-sions with on the order of 100 data points), there are far too many regions to consider an exhaustive search.  We therefore use a heuristic to find the most interesting region.  We imagine the region under consideration to be an amoeba moving within the unit box.  At each step in the search the amoeba either expands or contracts according to certain rules, and along the way we keep track of the most interesting $N$-region so far found, for each $N$.  The detailed rules for this heuristic are provided in Appendix~\ref{section:searchHeuristicDetails}.

\subsection{Steps 5 and 6:  Hypothetical similar experiments, \mbox{Part I}}
\label{section:HSEsI}

At this point in the algorithm the original events have been reduced to $N_{\rm data}$ values, each between 0 and 1:  the $p_N$ ($N = 1, \ldots, N_{\rm data}$) corresponding to the most interesting $N$-regions satisfying the imposed criteria.  To find the {\em most} interesting of these, we need a way of comparing regions of different size (different $N$).  An $N_1$-region $R_{N_1}$ with $p_{N_1}^{\rm data}$ is more interesting than an $N_2$-region $R_{N_2}$ with $p_{N_2}^{\rm data}$ if the fraction of hypothetical similar experiments in which $p_{N_1}^{\rm \hse} < p_{N_1}^{\rm data}$ is less than the fraction of hypothetical similar experiments in which $p_{N_2}^{\rm \hse} < p_{N_2}^{\rm data}$.

To make this comparison, we generate $N_{{\rm \hse}^1}$ hypothetical similar experiments.  Generating a hypothetical similar experiment involves pulling a random integer from \mbox{Eq.}~\ref{eqn:eqn7} for each background process $i$, sampling this number of events from the multidimensional background density $b(\vec{x})$, and then transforming these events into the unit box.  

For each \hse\ we compute a list of $p_N$, exactly as for the data set.  Each of the $N_{{\rm \hse}^1}$ hypothetical similar experiments consequently yields a list of $p_N$.  For each $N$, we now compare the $p_N$ we obtained in the data ($p_N^{\rm data}$) with the $p_N$'s we obtained in the \hse's ($p_N^{{\rm \hse}^1_i}$, where $i = 1, \ldots, N_{{\rm \hse}^1}$).  From these values we calculate $P_N$, the fraction of \hse's with $p_N^{{\rm \hse}^1} < p_N^{{\rm data}}$:
\begineq
P_N = \frac{1}{N_{{\rm \hse}^1}} \sum_{i=1}^{N_{{\rm \hse}^1}} { \Theta\left(p_N^{{\rm data}} - p_N^{{\rm \hse}^1_i} \right) },
\endeq
where $\Theta(x) = 0$ for $x < 0$, and $\Theta(x) = 1$ for $x \geq 0$.

The most interesting region in the sample is then the region for which $P_N$ is smallest.  We define $P = P_{N_{\rm min}}$, where $P_{N_{\rm min}}$ is the smallest of the $P_N$.

\subsection{Step 7:  Hypothetical similar experiments, \mbox{Part II}}

A question that remains to be answered is what fraction $\scriptP$ of hypothetical similar experiments would yield a $P$ less than the $P$ obtained in the data.  We calculate $\scriptP$ by running a second set of $N_{{\rm \hse}^2}$ hypothetical similar experiments, generated as described in the previous section.  (We have written \hse$^1$ above to refer to the first set of hypothetical similar experiments, used to determine the $P_N$, given a list of $p_N$; we write \hse$^2$ to refer to this second set of hypothetical similar experiments, used to determine $\scriptP$ from $P$.)  A second, independent set of \hse's is required to calculate an unbiased value for $\scriptP$.  The quantity $\scriptP$ is then given by
\begineq
\scriptP = \frac{1}{N_{{\rm \hse}^2}} \sum_{i=1}^{N_{{\rm \hse}^2}}{\Theta\left( P^{{\rm data}} - P^{{\rm \hse}^2_i} \right)}.
\endeq
This is the final measure of the degree of interest of the most interesting region.  Note that $\scriptP$ is a number between 0 and 1, that small values of $\scriptP$ indicate a sample containing an interesting region, that large values of $\scriptP$ indicate a sample containing no interesting region, and that $\scriptP$ can be described as the fraction of hypothetical similar experiments that yield a more interesting result than is observed in the data.  $\scriptP$ can be translated into units of standard deviations ($\scriptP_{[\sigma]}$) by solving the unit conversion equation
\begineq
\label{eqn:eqn13}
\scriptP = \frac{1}{\sqrt{2\pi}} \int_{\scriptP_{[\sigma]}}^{\infty} {e^{-t^2/2} \, dt}
\endeq
for $\scriptP_{[\sigma]}$.

\subsection{Interpretation of results}

In a general search for new phenomena, \Sherlock\ will be applied to $N_{\rm fs}$ different final states, resulting in $N_{\rm fs}$ different values for $\scriptP$.  The final step in the procedure is the combination of these results.  If no $\scriptP$ value is smaller than $\approx 0.01$ then a null result has been obtained, as no significant signal for new physics has been identified in the data.

If one or more of the $\scriptP$ values {\em is} particularly low, then we can surmise that the region(s) of excess corresponds either to a poorly modeled background or to possible evidence of new physics.  The algorithm has pointed out a region of excess ($\cal R$) and has quantified its significance ($\scriptP$).  The next step is to interpret this result.

Two issues related to this interpretation are combining results from many final states, and confirming a \Sherlock\ discovery.

\subsubsection{Combining the results of many final states}
\label{section:gothicP}

If one looks at many final states, one expects eventually to see a fairly small $\scriptP$, even if there really is no new physics in the data.  We therefore define a quantity $\gothicP$ to be the fraction of hypothetical similar {\em experimental runs}\footnote{In the phrase ``hypothetical similar experiment,'' ``experiment'' refers to the analysis of a single final state.  We use ``experimental runs'' in a similar way to refer to the analysis of a number of different final states.  Thus a hypothetical similar experimental run consists of $N_{\rm fs}$ different hypothetical similar experiments, one for each final state analyzed.} that yield a $\scriptP$ that is smaller than the smallest $\scriptP$ observed in the data.  Explicitly, given $N_{\rm fs}$ final states, with $\hat{b}_i$ background events expected in each, and $\scriptP_i$ calculated for each one, $\gothicP$ is given to good approximation by\footnote{Note that the naive expression $\gothicP = 1 - (1-\scriptP_{\min})^{N_{\rm fs}}$ is not correct, since this requires $\gothicP\rightarrow 1$ for $N_{\rm fs}\rightarrow\infty$, and there are indeed an infinite number of final states to examine.  The resolution of this paradox hinges on the fact that only an integral number of events can be observed in each final state, and therefore final states with $\hat{b}_i \ll 1$ contribute very little to the value of $\gothicP$.  This is correctly accounted for in the formulation given in Eq.~\ref{eqn:eqn16}.
}
\begineq
\label{eqn:eqn16}
\gothicP = 1 - \prod_{i=1}^{N_{\rm fs}}{ \sum_{j=0}^{n_i-1}{ \frac{e^{-\hat{b}_i} {\hat{b}_i}^j}{j!} }},
\endeq
where $n_i$ is the smallest integer satisfying
\begineq
\label{eqn:eqn17}
\sum_{j=n_i}^{\infty}{\frac{e^{-\hat{b}_i} {\hat{b}_i}^j}{j!}} \leq \scriptP_{\min} = \min_i{\scriptP_i}.
\endeq

\subsubsection{Confirmation}

An independent confirmation is desirable for any potential discovery, especially for an excess revealed by a data-driven search.  Such confirmation may come from an independent experiment, from the same experiment in a different but related final state, from an independent confirmation of the background estimate, or from the same experiment in the same final state using independent data.  In the last of these cases, a first sample can be presented to \Sherlock\ to uncover any hints of new physics, and the remaining sample can be subjected to a standard analysis in the region suggested by \Sherlock.  An excess in this region in the second sample helps to confirm a discrepancy between data and background.  If we see hints of new physics in the Run I data, for example, we will be able to predict where new physics might show itself in the upcoming run of the Fermilab Tevatron, Run II.

\begin{table}[htb]
\centering
\begin{tabular}{ll}
Final State	& Variables \\ \hline
$e\mu\met$ 	& $p_T^e$, $\met$ \\ 
$e\mu\met j$ 	& $p_T^e$, $\met$, $p_T^j$ \\
$e\mu\met j j$ 	& $p_T^e$, $\met$, $p_T^{j_2}$ \\ 
$e\mu\met jjj$	& $p_T^e$, $\met$, $p_T^{j_2}+p_T^{j_3}$ \\
\end{tabular}
\caption{The exclusive final states within $e\mu X$ for which events are seen in the data and the variables used for each of these final states.  The variables are selected using the prescription described in Sec.~\ref{section:SearchStrategy}.  Although all final states contain ``$e\mu\met$,'' no missing transverse energy cut has been applied explicitly; $\met$ is inferred from the presence of the muon, following Sec.~\ref{section:D0RunISearchStrategySupplement}.}
\label{tbl:tbl50}
\end{table}

\section{The ${\lowercase{e}}\mu X$ data set}
\label{section:theDataSet}

\def \EMMJZT {$3.0 \pm 0.8$} 
\widetext
\begin{table*}[bht]
\centering
\small
\begin{tabular}{ccccccc}
Data set & Fakes & $Z\rightarrow\tau\tau$ & $\gamma^*\rightarrow\tau\tau$ & $WW$ & \ttbar & Total\\ \hline
$e\mu\met$ & \EMMFA & \EMMZT & \EMMDY & \EMMWW & \EMMTT & \EMMTOT\\
$e\mu\met j$ & \EMMJFA & \EMMJZT & \EMMJDY & \EMMJWW & \EMMJTT & \EMMJTOT \\
$e\mu\met jj$ & \EMMJJFA & \EMMJJZT & \EMMJJDY & \EMMJJWW & \EMMJJTT & \EMMJJTOT \\
$e\mu\met jjj$ &\EMMJJJFA & \EMMJJJZT & \EMMJJJDY & \EMMJJJWW & \EMMJJJTT & \EMMJJJTOT \\ \hline
$e\mu X$ & \EMXFA & \EMXZT & \EMXDY & \EMXWW & \EMXTT & \EMXTOT \\
\end{tabular}
\caption{The number of expected background events for the populated final states within $e\mu X$. The errors on $e\mu X$ are smaller than on the sum of the individual background contributions obtained from Monte Carlo because of an uncertainty on the number of extra jets arising from initial and final state radiation in the exclusive channels.}
\label{expectedBackgrounds1}
\end{table*}
\narrowtext

As mentioned in Sec.~\ref{section:introduction}, we have applied the \Sherlock\ method to D\O\ data containing one or more electrons and one or more muons.  We use a data set corresponding to 108.3$\pm$5.7 pb$^{-1}$ of integrated luminosity, collected between 1992 and 1996 at the Fermilab Tevatron with the D\O\ detector.  The data set and basic selection criteria are identical to those used in the published $t\bar{t}$ cross section analysis for the dilepton channels~\cite{topCrossSection}.  Specifically, we apply global cleanup cuts and select events containing
\begin{itemize}
\item{one or more high $p_T$ ($p_T>15$~GeV) isolated electrons, and}
\item{one or more high $p_T$ ($p_T>15$~GeV) isolated muons,}
\end{itemize}
with object definitions given in Sec.~\ref{section:D0RunISearchStrategySupplement}.

The dominant standard model and instrumental backgrounds to this data set are
\begin{itemize}
\item{top quark pair production with $t\rightarrow W b$, and with both $W$ bosons decaying leptonically, one to $e\nu$ (or to $\tau\nu\rightarrow e\nu\nu\nu$) and one to $\mu\nu$ (or to $\tau\nu\rightarrow \mu\nu\nu\nu$),}
\item{$W$ boson pair production with both $W$ bosons decaying leptonically, one to $e\nu$ (or to $\tau\nu\rightarrow e\nu\nu\nu$) and one to $\mu\nu$ (or to $\tau\nu\rightarrow \mu\nu\nu\nu$),}
\item{$Z/\gamma^*\rightarrow \tau\tau \rightarrow e\mu\nu\nu\nu\nu$, and}
\item{instrumental (``fakes''):  $W$ production with the $W$ boson decaying to $\mu\nu$ and a radiated jet or photon being mistaken for an electron, or $b\bar{b}/c\bar{c}$ production with one heavy quark producing an isolated muon and the other a false electron~\cite{searchForTopPRD}}.
\end{itemize}
A sample of 100,000 $t\bar{t}\rightarrow$ dilepton events was generated using {\small HERWIG}~\cite{HERWIG}, and a $WW$ sample of equal size was generated using {\small PYTHIA}~\cite{PYTHIA}.  We generated $\gamma^*\rightarrow \tau\tau \rightarrow e\mu\nu\nu\nu\nu$ (Drell-Yan) events using {\small PYTHIA} and $Z\rightarrow \tau\tau \rightarrow e\mu\nu\nu\nu\nu$ events using {\small ISAJET}~\cite{ISAJET}.  The Drell-Yan cross section is normalized as in Ref.~\cite{McKinley}.  The cross section for $Z\rightarrow \tau\tau$ is taken to be equal to the published D\O\ $Z\rightarrow ee$ cross section~\cite{ZeeCrossSection}; the top quark production cross section is taken from Ref.~\cite{Laenen}; and the $WW$ cross section is taken from Ref.~\cite{Baer}.  The $t\bar{t}$, $WW$, and $Z/\gamma^*$ Monte Carlo events all were processed through {\small GEANT}~\cite{GEANT} and the D\O\ reconstruction software.  The number and distributions of events containing fake electrons are taken from data, using a sample of events satisfying ``bad'' electron identification criteria~\cite{directTopMassMeasurement}.

We break $e\mu X$ into exclusive data sets, and determine which variables to consider in each set using the prescription given in Sec.~\ref{section:SearchStrategy}.  The exclusive final states within $e\mu X$ that are populated with events in the data are listed in Table~\ref{tbl:tbl50}.  The number of events expected for the various samples and data sets in the populated final \mbox{states} are given in Table~\ref{expectedBackgrounds1}; the number of expected background events in all unpopulated final states in which the number of expected background events is $> 0.001$ are listed in Table~\ref{expectedBackgrounds2}.  The dominant sources of systematic error are given in Table~\ref{Errors Table}.  

\begin{table}[htb]
\centering
\small
\begin{tabular}{lc}
Final State & Background expected \\ \hline
$e\mu\met jjjj$ & $0.3 \pm 0.15$ \\
$ee\mu\met$ & $0.10 \pm 0.05$ \\
$e\mu\mu$ & $0.04 \pm 0.02$ \\
$e\mu\met\gamma$ & $0.06 \pm 0.03$ \\
\end{tabular}
\caption{The number of expected background events for the unpopulated final states within $e\mu X$.  The expected number of events in final states with additional jets is obtained from those listed in the table by dividing by five for each jet.  These are all rough estimates, and a large systematic error has been assigned accordingly.  Since no events are seen in any of these final states, the background estimates shown here are used solely in the calculation of $\gothicP$ for all $e\mu X$ channels.}
\label{expectedBackgrounds2}
\end{table}

\begin{table}[htb]
\centering
\begin{tabular}{lc}
Source & Error \\ \hline
Trigger and lepton identification efficiencies & 12\% \\ 
$P(j\rightarrow$``$e$''$)$ & 7\% \\ 
Multiple Interactions & 7\% \\ 
Luminosity  & 5.3\% \\ 
$\sigma$(\ttbar$\rightarrow e\mu X$) & 12\% \\ 
$\sigma(Z\rightarrow\tau\tau\rightarrow e\mu X)$   & 10\% \\ 
$\sigma(WW\rightarrow e\mu X)$     & 10\% \\ 
$\sigma(\gamma^*\rightarrow\tau\tau\rightarrow e\mu X$)  & 17\% \\ 
Jet modeling  & 20\% \\ 
\end{tabular}
\caption{Sources of systematic uncertainty on the number of expected background events in the final states $e\mu\met$, $e\mu\met j$, $e\mu\met jj$, and $e\mu\met jjj$.  $P(j\rightarrow$``$e$''$)$ denotes the probability that a jet will be reconstructed as an electron.  ``Jet modeling'' includes systematic uncertainties in jet production in {\small PYTHIA} and {\small HERWIG} in addition to jet identification and energy scale uncertainties.}
\label{Errors Table}
\end{table}

\section{Sensitivity}
\label{section:sensitivity}

We choose to consider the $e\mu X$ final state first because it contains backgrounds of mass scale comparable to that expected of the physics responsible for electroweak symmetry breaking.  Top quark pair production ($q\bar{q}\rightarrow t\bar{t} \rightarrow W^+ W^- b\bar{b}$) and $W$ boson pair production are excellent examples of the type of physics that we would expect the algorithm to find.

Before examining the data, we decided to impose the requirements of AntiCornerSphere and Isolation (see Appendix~\ref{section:RegionCriteriaAppendix}) on the regions that \Sherlock\ is allowed to consider.  The reason for this choice is that, in addition to allowing only ``reasonable'' regions, it allows the search to be parameterized essentially by a single variable --- the distance between each region and the lower left-hand corner of the unit box.  We felt this would aid the interpretation of the results from this initial application of the method.  

We test the sensitivity in two phases, keeping in mind that nothing in the algorithm has been ``tuned'' to finding $WW$ and $t\bar{t}$ in this sample.  We first consider the background to comprise fakes and $Z/\gamma^* \rightarrow\tau\tau$ only, to see if we can ``discover'' either $WW$ or $t\bar{t}$.  We then consider the background to comprise fakes, $Z/\gamma^* \rightarrow\tau\tau$, and $WW$, to see whether we can ``discover'' $t\bar{t}$.  We apply the full search strategy and algorithm in both cases, first (in this section) on an ensemble of mock samples, and then (in Sec.~\ref{section:Results}) on the data.

\subsection{Search for $WW$ and $t\bar{t}$ in mock samples}
\label{section:findWWandttbar}

In this section we provide results from \Sherlock\ for the case in which $Z/\gamma^*\rightarrow\tau\tau$ and fakes are included in the background estimates and the signal from $WW$ and $t\bar{t}$ is ``unknown.''  We apply the prescription to the exclusive $e\mu X$ final states listed in Table~\ref{tbl:tbl50}.

Figure~\ref{fig:exer_wwttbar_backonly} shows distributions of $\scriptP$ for mock samples containing only $Z/\gamma^*\rightarrow\tau\tau$ and fakes, where the mock events are pulled randomly from their parent distributions and the numbers of events are allowed to vary within systematic and statistical errors.  The distributions are uniform in the interval $[0,1]$, as expected, becoming appropriately discretized in the low statistics limit.  (When the number of expected background events $\hat{b} \ltapprox 1$, as in Fig.~\ref{fig:exer_wwttbar_backonly}(d), it can happen that zero or one events are observed.  If zero events are observed then $\scriptP=1$, since all hypothetical similar experiments yield a result as interesting or more interesting than an empty sample.  If one event is observed then there is only one region for \Sherlock\ to consider, and $\scriptP$ is simply the probability for $\hat{b}\pm\delta\hat{b}$ to fluctuate up to exactly one event.  In Fig.~\ref{fig:exer_wwttbar_backonly}(d), for example, the spike at $\scriptP=1$ contains 62\% of the mock experiments, since this is the probability for $0.5\pm0.2$ to fluctuate to zero events; the second spike is located at $\scriptP = 0.38$ and contains 28\% of the mock experiments, since this is the probability for $0.5\pm0.2$ to fluctuate to exactly one event.  Similar but less pronounced behavior is seen in Fig.~\ref{fig:exer_wwttbar_backonly}(c).) Figure~\ref{fig:exer_wwttbar_signal} shows distributions of $\scriptP$ when the mock samples contain $WW$ and $t\bar{t}$ in addition to the background in Fig.~\ref{fig:exer_wwttbar_backonly}.  \mbox{Again}, the number of events from each process is allowed to vary within statistical and systematic error.  Figure~\ref{fig:exer_wwttbar_signal} shows that we can indeed find $t\bar{t}$ and/or $WW$ much of the time.  Figure~\ref{fig:exer_wwttbar_twidpsig} shows $\gothicP$ computed for these samples.  In over 50\% of these samples we find $\gothicP_{[\sigma]}$ to correspond to more than two standard deviations.

{\dofig {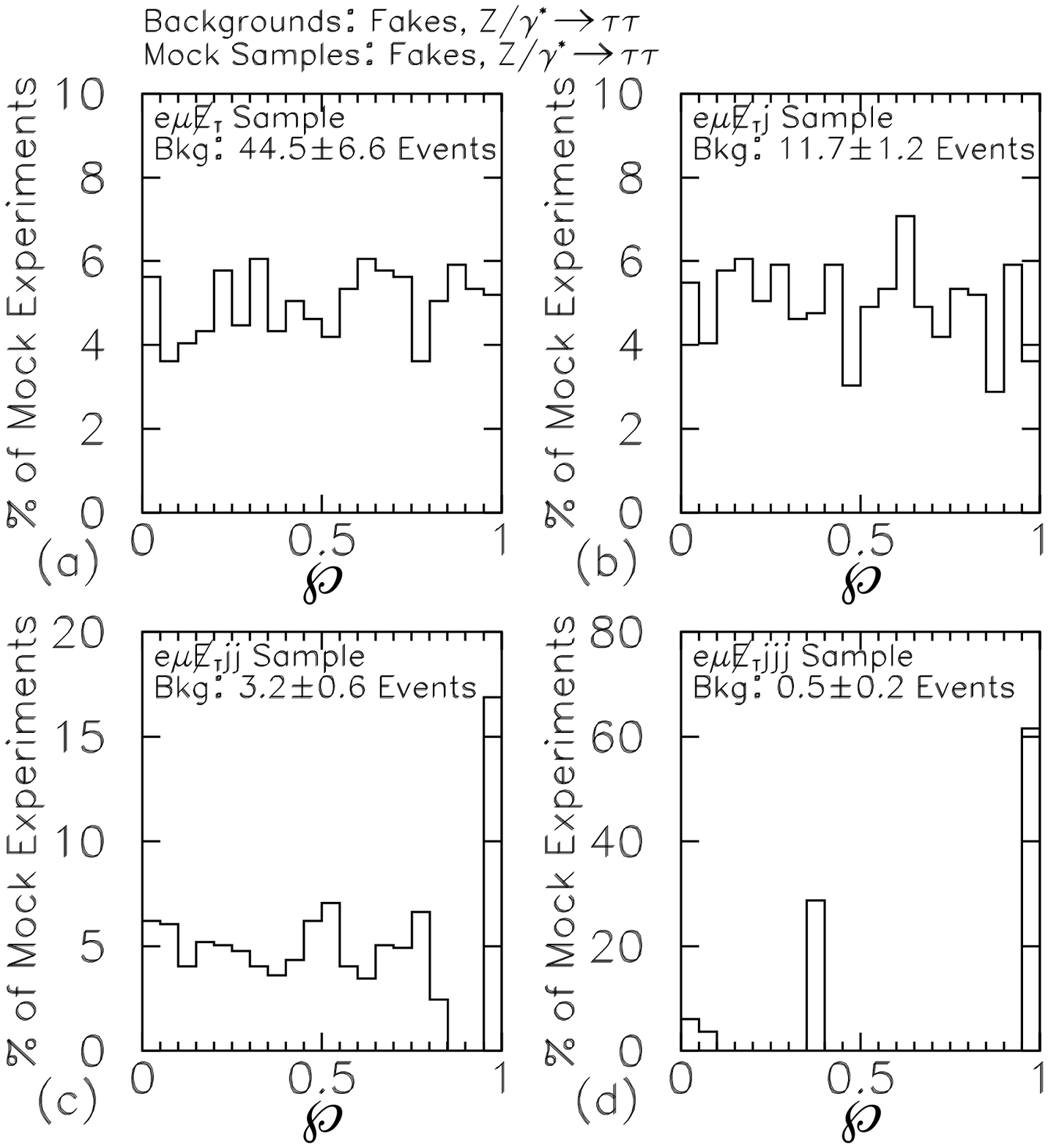} {3.5in} {Distributions of $\scriptP$ for the four exclusive final states (a) $e\mu\met$, (b) $e\mu\met j$, (c) $e\mu\met jj$, and (d) $e\mu\met jjj$.  The background includes only $Z/\gamma^*\rightarrow\tau\tau$ and fakes, and the mock samples making up these distributions also contain only these two sources.  As expected, $\scriptP$ is uniform in the interval $[0,1]$ for those final states in which the expected number of background events $\hat{b} \gg 1$, and shows discrete behavior for $\hat{b}$ \protect\raisebox{-0.6ex}{$\stackrel{<}{\sim}$} 1.} {fig:exer_wwttbar_backonly}}
{\dofig {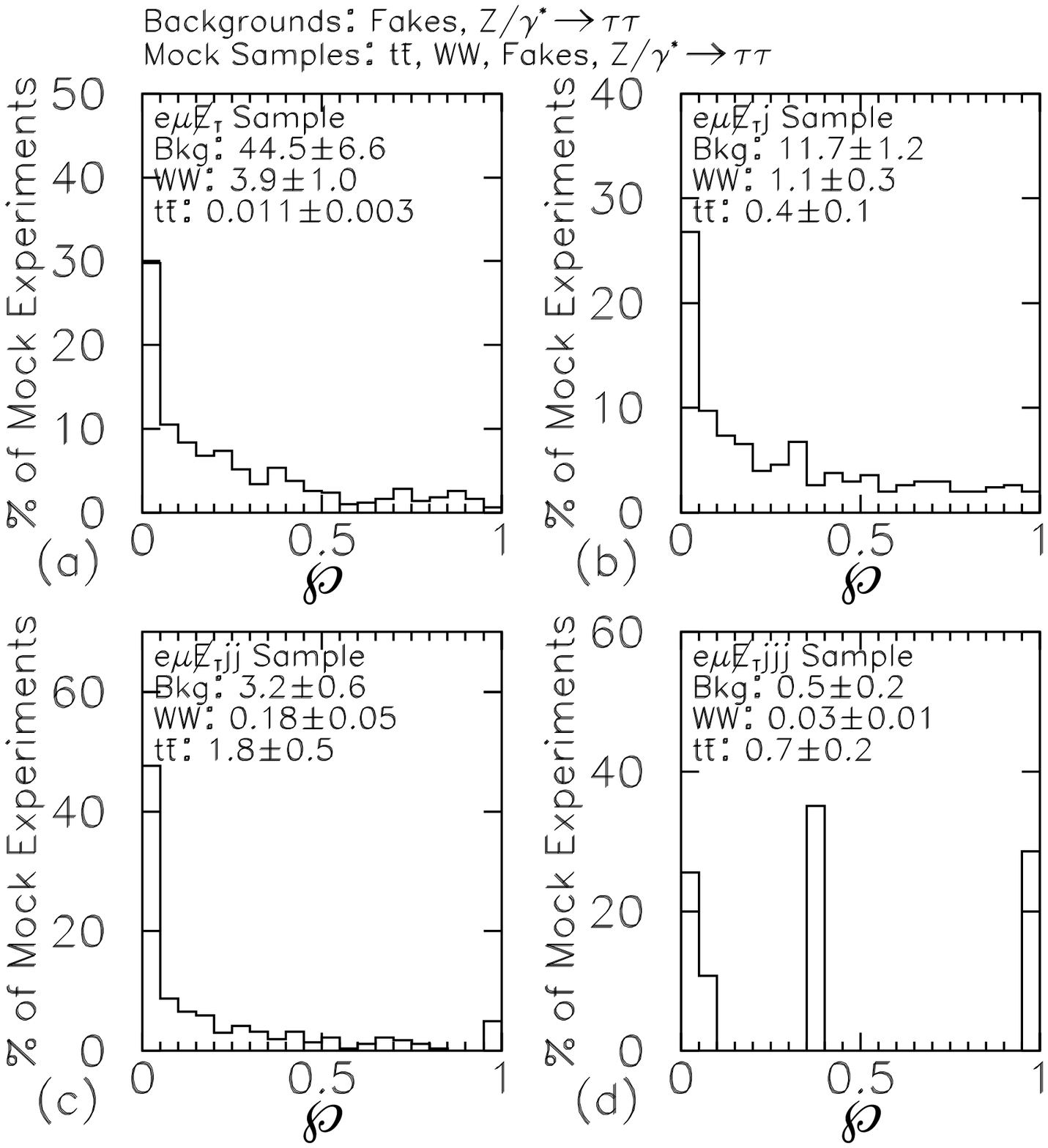} {3.5in} {Distributions of $\scriptP$ for the four exclusive final \mbox{states} (a) $e\mu\met$, (b) $e\mu\met j$, (c) $e\mu\met jj$, and (d) $e\mu\met jjj$.  The background includes only $Z/\gamma^*\rightarrow\tau\tau$ and fakes.  The mock samples for these distributions contain $WW$ and $t\bar{t}$ in addition to $Z/\gamma^*\rightarrow\tau\tau$ and fakes.  The extent to which these distributions peak at small $\scriptP$ can be taken as a measure of \Sherlock's ability to find $WW$ or $t\bar{t}$ if we had no knowledge of either final state.  The presence of $WW$ in $e\mu\met$ causes the trend toward small values in (a); the presence of $t\bar{t}$ causes the trend toward small values in (c) and (d); and a combination of $WW$ and $t\bar{t}$ causes the signal seen in (b).} {fig:exer_wwttbar_signal}}
{\dofig {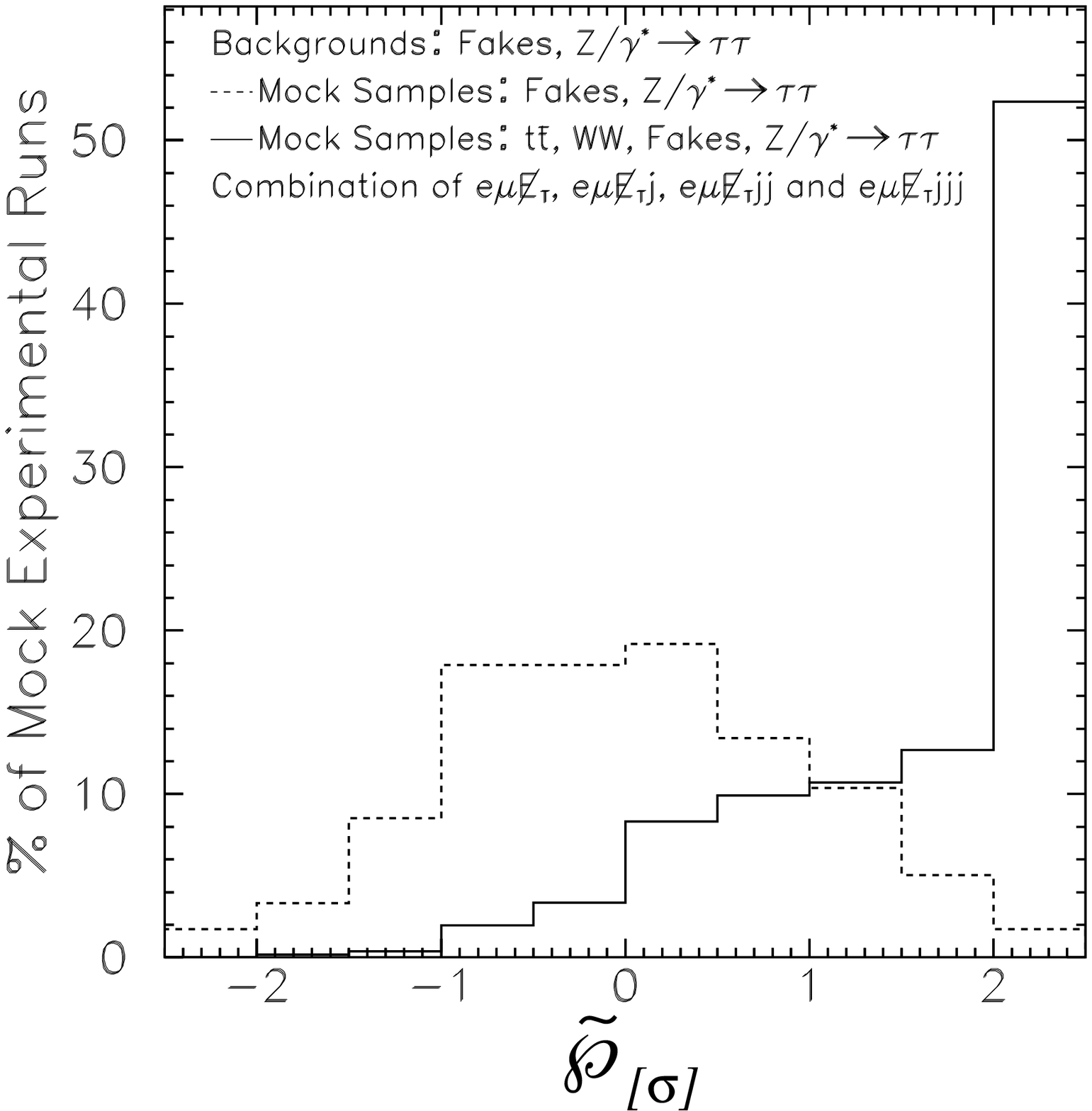} {3.5in} {Distribution of $\gothicP_{[\sigma]}$ from combining the four exclusive final states $e\mu\met$, $e\mu\met j$, $e\mu\met jj$, and $e\mu\met jjj$.  The background includes only $Z/\gamma^*\rightarrow\tau\tau$ and fakes.  The mock samples making up the distribution shown as the solid line contain $WW$ and $t\bar{t}$ in addition to $Z/\gamma^*\rightarrow\tau\tau$ and fakes, and correspond to Fig.~\ref{fig:exer_wwttbar_signal}; the mock samples making up the distribution shown as the dashed line contain only $Z/\gamma^*\rightarrow\tau\tau$ and fakes, and correspond to Fig.~\ref{fig:exer_wwttbar_backonly}.  All samples with $\gothicP_{[\sigma]}>2.0$ appear in the rightmost bin.  The fact that $\gothicP_{[\sigma]}>2.0$ in 50\% of the mock samples can be taken as a measure of \Sherlock's sensitivity to finding $WW$ and $t\bar{t}$ if we had no knowledge of the existence of the top quark or the possibility of $W$ boson pair production.} {fig:exer_wwttbar_twidpsig}}

\subsection{Search for $t\bar{t}$ in mock samples}
\label{section:findttbar}

In this section we provide results for the case in which $Z/\gamma^*\rightarrow\tau\tau$, fakes, and $WW$ are all included in the background estimate, and $t\bar{t}$ is the ``unknown'' signal.  We again apply the prescription to the exclusive final states listed in Table~\ref{tbl:tbl50}.

Figure~\ref{fig:exer_ttbar_backonly} shows distributions of $\scriptP$ for mock samples containing $Z/\gamma^*\rightarrow\tau\tau$, fakes, and $WW$, where the mock events are pulled randomly from their parent distributions, and the numbers of events are allowed to vary within systematic and statistical errors.  As found in the previous section, the distributions are uniform in the interval $[0,1]$, becoming appropriately discretized when the expected number of background events becomes  \raisebox{-0.6ex}{$\stackrel{<}{\sim}$} 1.  Figure~\ref{fig:exer_ttbar_signal} shows distributions of $\scriptP$ when the mock samples contain $t\bar{t}$ in addition to $Z/\gamma^*\rightarrow\tau\tau$, fakes, and $WW$.  Again, the number of events from each process is allowed to vary within statistical and systematic errors.  The distributions in Figs.~\ref{fig:exer_ttbar_signal}(c) and (d) show that we can indeed find $t\bar{t}$ much of the time.  Figure~\ref{fig:exer_ttbar_twidpsig} shows that the distribution of $\gothicP_{[\sigma]}$ is approximately a Gaussian centered at zero of width unity for the case where the background and data both contain $Z/\gamma^*\rightarrow\tau\tau$, fakes, and $WW$ production, and is peaked in the bin above $2.0$ for the same background when the data include $t\bar{t}$.

{\dofig {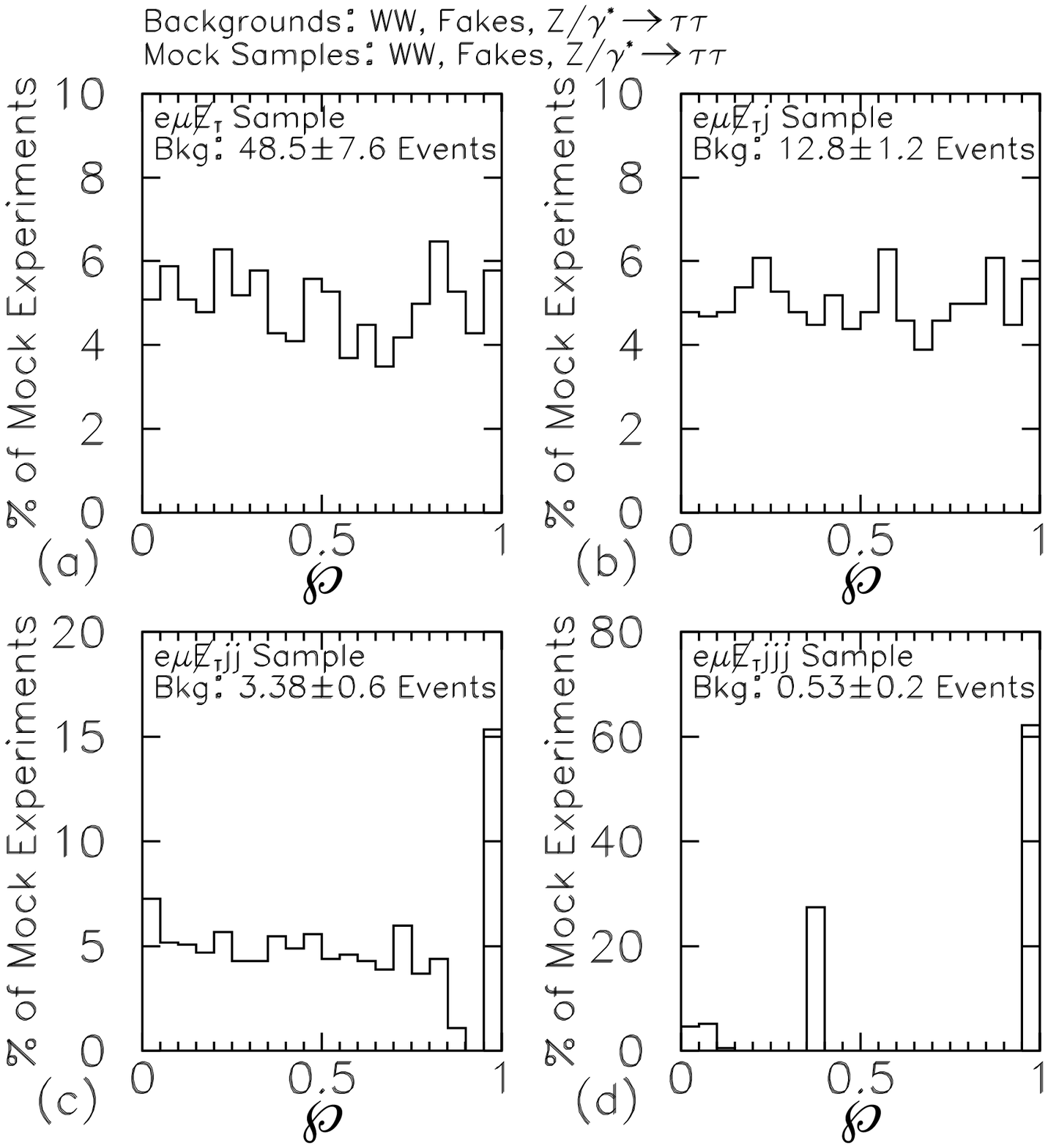} {3.5in} {Distributions of $\scriptP$ for the four exclusive final states (a) $e\mu\met$, (b) $e\mu\met j$, (c) $e\mu\met jj$, and (d) $e\mu\met jjj$.  The background includes $Z/\gamma^*\rightarrow\tau\tau$, fakes, and $WW$, and the mock samples making up these distributions also contain these three sources.  As expected, $\scriptP$ is uniform in the interval $[0,1]$ for those final states in which the expected number of background events $\hat{b} \gg 1$, and shows discrete behavior when $\hat{b}$ \protect\raisebox{-0.6ex}{$\stackrel{<}{\sim}$} 1.} {fig:exer_ttbar_backonly}}
{\dofig {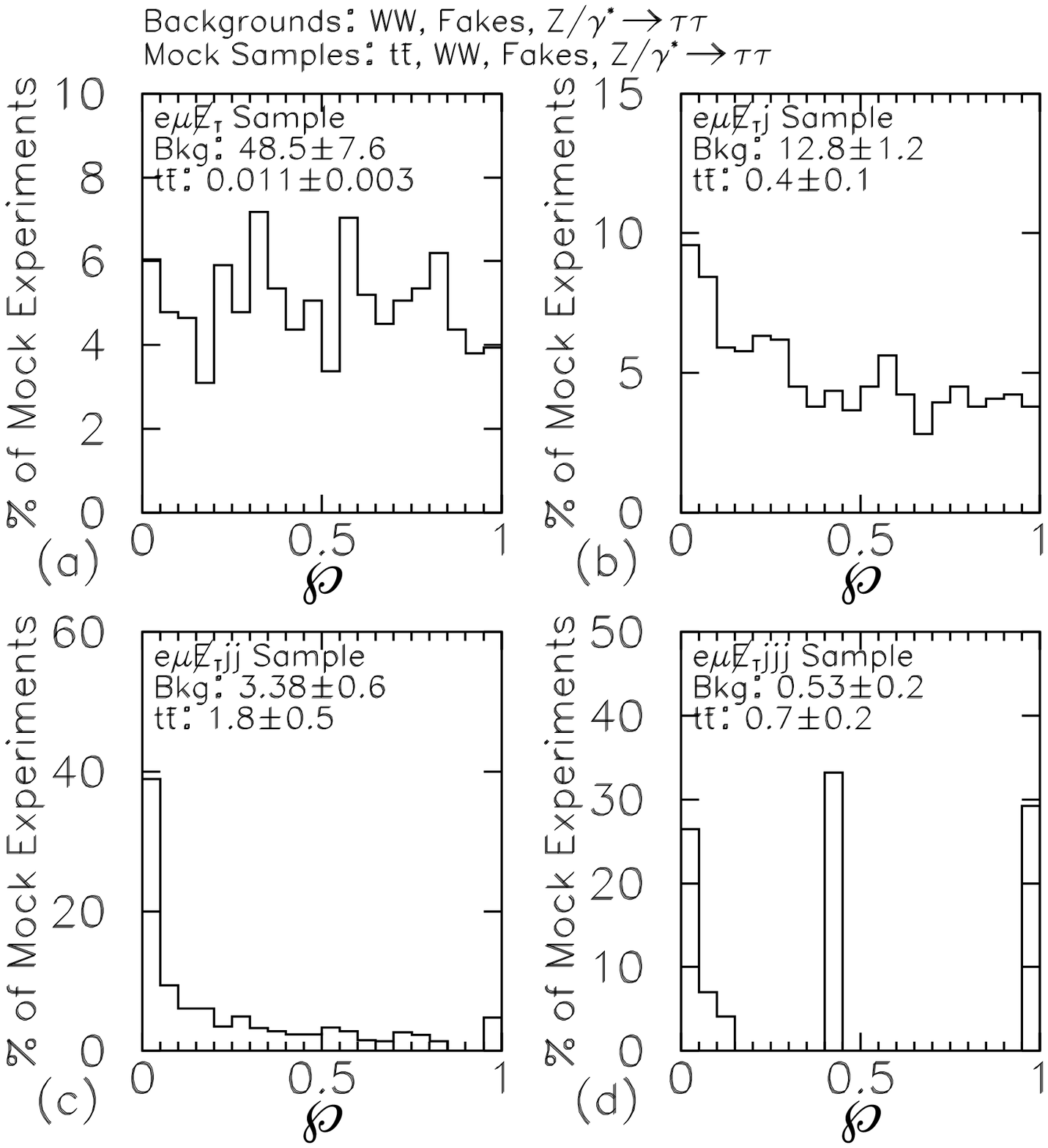} {3.5in} {Distributions of $\scriptP$ for the four exclusive final \mbox{states} (a) $e\mu\met$, (b) $e\mu\met j$, (c) $e\mu\met jj$, and (d) $e\mu\met jjj$.  The background includes $Z/\gamma^*\rightarrow\tau\tau$,  fakes, and $WW$.  The mock samples for these distributions contain $t\bar{t}$ in addition to $Z/\gamma^*\rightarrow\tau\tau$, fakes, and $WW$.  The extent to which these distributions peak at small $\scriptP$ can be taken as a measure of \Sherlock's sensitivity to finding $t\bar{t}$ if we had no knowledge of the top quark's existence or characteristics.  Note that $\scriptP$ is flat in $e\mu\met$, where the expected number of top quark events is negligible, peaks slightly toward small values in $e\mu\met j$, and shows a marked low peak in $e\mu\met j j$ and $e\mu\met jjj$.} {fig:exer_ttbar_signal}}
{\dofig {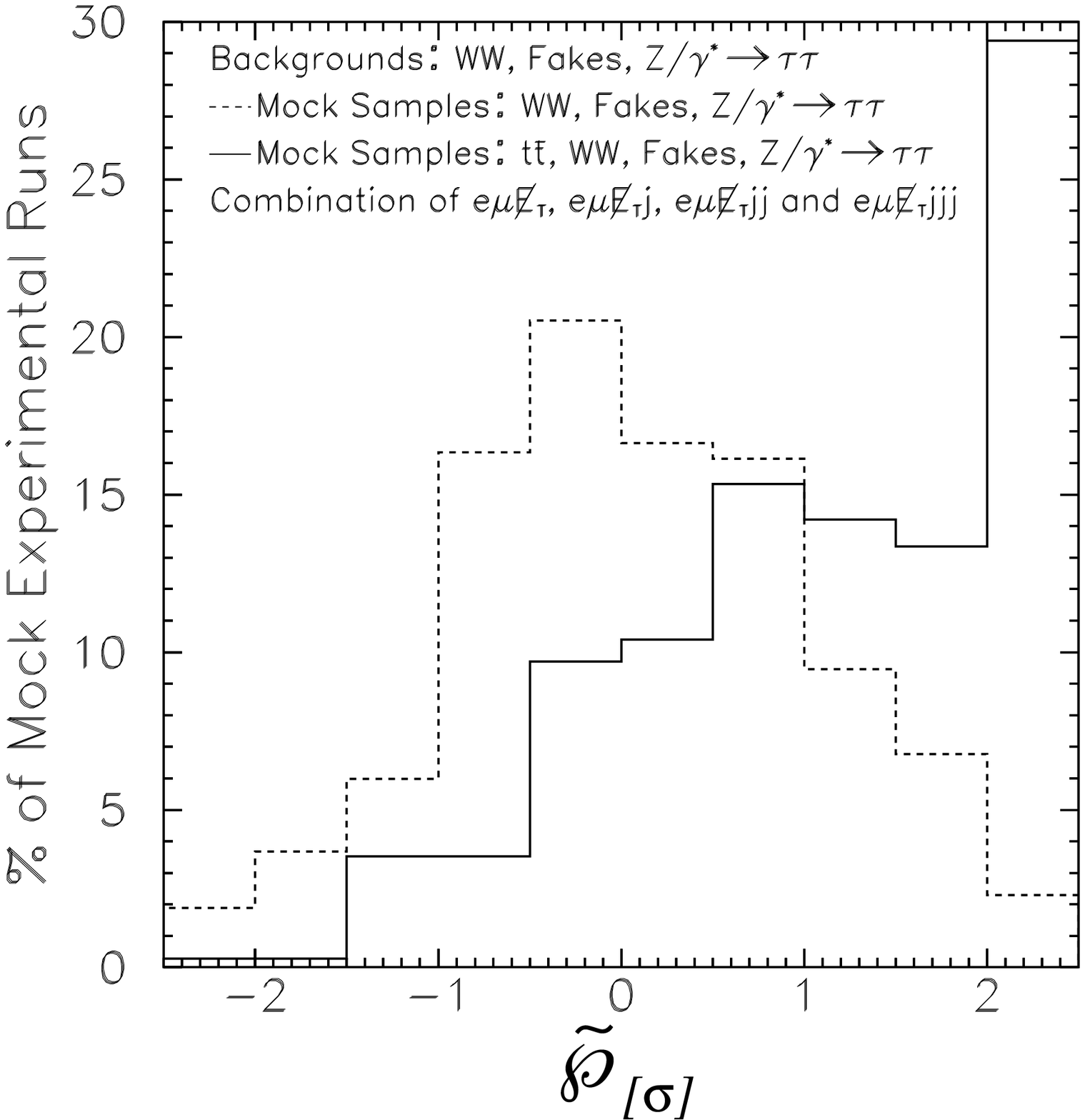} {3.5in} {Distribution of $\gothicP_{[\sigma]}$ from combining the four exclusive final states $e\mu\met$, $e\mu\met j$, $e\mu\met jj$, and $e\mu\met jjj$.  The background includes $Z/\gamma^*\rightarrow\tau\tau$, fakes, and $WW$.  The mock samples making up the distribution shown as the solid line contain $t\bar{t}$ in addition to $Z/\gamma^*\rightarrow\tau\tau$, fakes, and $WW$, corresponding to Fig.~\ref{fig:exer_ttbar_signal}; the mock samples making up the distribution shown as the dashed line contain only $Z/\gamma^*\rightarrow\tau\tau$, fakes, and $WW$, and correspond to Fig.~\ref{fig:exer_ttbar_backonly}.  All samples with $\gothicP_{[\sigma]}>2.0$ appear in the rightmost bin.  The fact that $\gothicP_{[\sigma]}>2.0$ in over 25\% of the mock samples can be taken as a measure of \Sherlock's sensitivity to finding $t\bar{t}$ if we had no knowledge of the top quark's existence or characteristics.} {fig:exer_ttbar_twidpsig}}

\subsection{New high $p_T$ physics}
\label{section:RoughSensitivityScale}

We have shown in Secs.~\ref{section:findWWandttbar} and~\ref{section:findttbar} that the \Sherlock\ prescription and algorithm correctly finds nothing when there is nothing to be found, while exhibiting sensitivity to the expected presence of $WW$ and $t\bar{t}$ in the $e\mu X$ sample.  \Sherlock's performance on this ``typical'' new physics signal is encouraging, and may be taken as some measure of the sensitivity of this method to the great variety of new high $p_T$ physics that it has been designed to find.  Making a more general claim regarding \Sherlock's sensitivity to the presence of new physics is difficult, since the sensitivity obviously varies with the characteristics of each candidate theory.  

That being said, we can provide a rough estimate of \Sherlock's sensitivity to new high $p_T$ physics with the following argument.  We have seen that we are sensitive to $WW$ and $t\bar{t}$ pair production in a data sample corresponding to an integrated luminosity of $\approx 100$~pb$^{-1}$.  These events tend to fall in the region $p_T^e>40$~GeV, $\met>40$~GeV, and $\sum'{p_T^j}>40$~GeV (if there are any jets at all).  The probability that any true $e\mu X$ event produced will make it into the final sample is about 15\% due to the absence of complete hermeticity of the D\O\ detector, inefficiencies in the detection of electrons and muons, and kinematic acceptance.  We can therefore state that we are as sensitive to new high $p_T$ physics as we were to the roughly eight $WW$ and $t\bar{t}$ events in our mock samples if the new physics is distributed relative to all standard model backgrounds as $WW$ and $t\bar{t}$ are distributed relative to backgrounds from $Z/\gamma^*\rightarrow\tau\tau$ and fakes alone, and if its production cross section $\times$ branching ratio into this final state is \raisebox{-0.6ex}{$\stackrel{>}{\sim}$} $8/(0.15\times 100~{\rm pb}^{-1}) \approx 600$~fb.  Readers who are interested in a possible signal with a different relative distribution, or who prefer a more rigorous definition of ``sensitivity,'' should adjust this cross section accordingly.


\section{Results}
\label{section:Results}

In the previous section we studied what can be expected when \Sherlock\ is applied to $e\mu X$ mock samples.  In this section we confront \Sherlock\ with data.  We observe 39 events in the $e\mu\met$ final state, 13 events in $e\mu\met j$, 5 events in $e\mu\met jj$, and a single event in $e\mu\met jjj$, in good agreement with the expected background in Table~\ref{expectedBackgrounds1}.  We proceed by first removing both $WW$ and $t\bar{t}$ from the background estimates, and next by removing only $t\bar{t}$, to search for evidence of these processes in the data.  Finally, we include all standard model processes in the background estimates and search for evidence of new physics.

\subsection{Search for $WW$ and $t\bar{t}$ in data}

The results of applying \Sherlock\ to D\O\ data with only $Z/\gamma^*\rightarrow\tau\tau$ and fakes in the background estimate are shown in Table~\ref{tbl:tbl01} and Fig.~\ref{fig:data_mapping_wwttbar}.  \Sherlock\ finds indications of an excess in the $e\mu\met$ and $e\mu\met jj$ states, presumably reflecting the presence of $WW$ and $t\bar{t}$, respectively.  The results for the $e\mu\met j$ and $e\mu\met jjj$ final states are consistent with the results in Fig.~\ref{fig:exer_wwttbar_signal}.  Defining $r'$ as the distance of the data point from $(0,0,0)$ in the unit box (transformed so that the background is distributed uniformly in the interval $[0,1]$), the top candidate events from D\O's recent analysis~\cite{Harpreet} are the three events with largest $r'$ in the $e\mu\met jj$ sample and the single event in the $e\mu\met jjj$ sample, shown in Fig.~\ref{fig:data_mapping_wwttbar}.  The presence of the $WW$ signal can be inferred from the events designated interesting in the $e\mu\met$ final state.

\begin{table}[htb]
\centering
\begin{tabular}{cc}
Data set  	& $\scriptP$ 	\\ \hline
$e\mu\met$   	& 0.008		\\
$e\mu\met j$  	& 0.34		\\
$e\mu\met jj$ 	& 0.01		\\
$e\mu\met jjj$  & 0.38		\\ \hline
\raisebox{-.6ex}{$\gothicP$}	& 0.03 \\ 
\end{tabular}
\caption{Summary of results on the $e\mu\met$, $e\mu\met j$, $e\mu\met jj$, and $e\mu\met jjj$ channels when $WW$ and $t\bar{t}$ are not included in the background.  \Sherlock\ identifies a region of excess in the $e\mu\met$ and $e\mu\met jj$ final states, presumably indicating the presence of $WW$ and $t\bar{t}$ in the data.  In units of standard deviation, $\gothicP_{[\sigma]} = 1.9$.}
\label{tbl:tbl01}
\end{table}

{\dofig {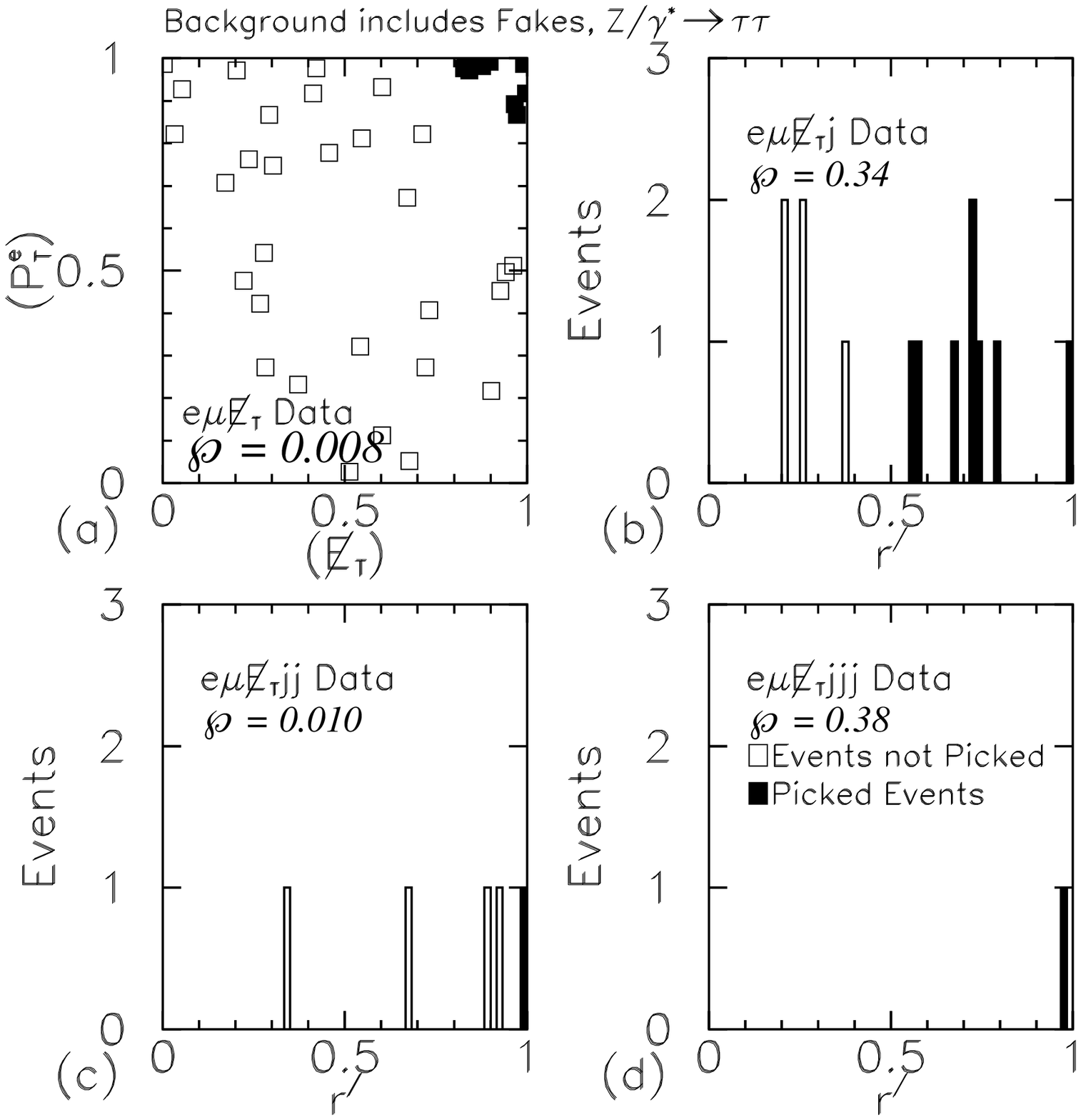} {3.5in} {Positions of data points following the transformation of the background from fake and $Z/\gamma^*$ sources in the space of variables in Table~\ref{tbl:VariableRules} to a uniform distribution in the unit box.  The darkened points define the region \Sherlock\ found most interesting.  The axes of the unit box in (a) are suggestively labeled $(p_T^e)$ and $(\met)$; each is a function of both $p_T^e$ and $\met$, but $(p_T^e)$ depends more strongly on $p_T^e$, while $(\met)$ more closely tracks $\met$.  $r'$ is the distance of the data point from $(0,0,0)$ (the ``lower left-hand corner'' of the unit box), transformed so that the background is distributed uniformly in the interval $[0,1]$.  The interesting regions in the $e\mu\met$ and $e\mu\met jj$ samples presumably indicate the presence of $WW$ signal in $e\mu\met$ and of $t\bar{t}$ signal in $e\mu\met jj$.  We find $\gothicP=0.03$ ($\gothicP_{[\sigma]} = 1.9$).} {fig:data_mapping_wwttbar}}

\subsection{Search for $t\bar{t}$ in data}

The results of applying \Sherlock\ to the data with $Z/\gamma^*\rightarrow\tau\tau$, fakes, and $WW$ included in the background estimate are shown in Table~\ref{tbl:tbl02} and Fig.~\ref{fig:data_mapping_ttbar}.  \Sherlock\ finds an indication of excess in the $e\mu\met jj$ events, presumably indicating the presence of $t\bar{t}$.  The results for the $e\mu\met$, $e\mu\met j$, and $e\mu\met jjj$ final states are consistent with the results in Fig.~\ref{fig:exer_ttbar_signal}.  The $t\bar{t}$ candidates from D\O's recent analysis~\cite{Harpreet} are the three events with largest $r'$ in the $e\mu\met jj$ sample, and the single event in the $e\mu\met jjj$ sample, shown in Fig.~\ref{fig:data_mapping_ttbar}.

\begin{table}[htb]
\centering
\begin{tabular}{cc}
Data set  	& $\scriptP$ 	\\ \hline
$e\mu\met$   	& 0.16		\\
$e\mu\met j$  	& 0.45		\\
$e\mu\met jj$ 	& 0.03		\\
$e\mu\met jjj$  & 0.41		\\ \hline
\raisebox{-.6ex}{$\gothicP$}	& 0.11  	\\ 
\end{tabular}
\caption{Summary of results on the $e\mu\met$, $e\mu\met j$, $e\mu\met jj$, and $e\mu\met jjj$ channels when $t\bar{t}$ production is not included in the background.  \Sherlock\ identifies a region of excess in the $e\mu\met jj$ final state, presumably indicating the presence of $t\bar{t}$ in the data.    In units of standard deviation, $\gothicP_{[\sigma]} = 1.2$.}
\label{tbl:tbl02}
\end{table}

{\dofig {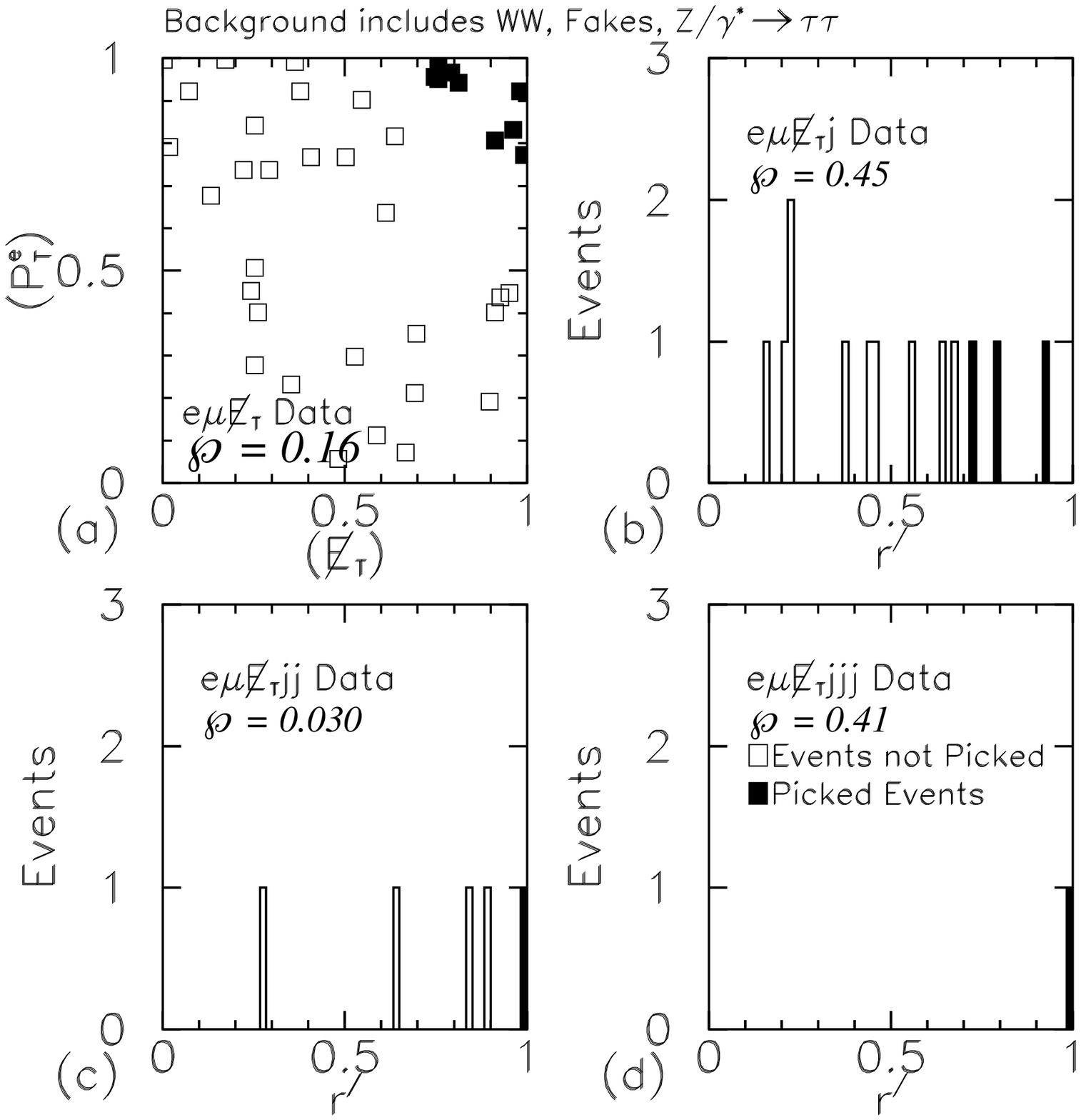} {3.5in} {Positions of data points following the transformation of the background from the three sources $Z/\gamma^*\rightarrow\tau\tau$, fakes, and $WW$ in the space of variables in Table~\ref{tbl:VariableRules} to a uniform distribution in the unit box.  The darkened points define the region \Sherlock\ found most interesting.  The interesting region in the $e\mu\met jj$ sample presumably indicates the presence of $t\bar{t}$.  We find $\gothicP=0.11$ ($\gothicP_{[\sigma]} = 1.2$).} {fig:data_mapping_ttbar}}

A comparison of this result with one obtained using a dedicated top quark search illustrates an important difference between \Sherlock's result and the result from a dedicated search.  D\O\ announced its discovery of the top quark~\cite{topQuarkObservation} in 1995 with 50~pb$^{-1}$ of integrated luminosity upon observing 17 events with an expected background of $3.8\pm0.6$ events, a $4.6\sigma$ ``effect,'' in the combined dilepton and single-lepton decay channels.  In the $e\mu$ channel alone, two events were seen with an expected background of $0.12 \pm 0.03$ events.  The probability of $0.12 \pm 0.03$ events fluctuating up to or above two events is $0.007$, corresponding to a $2.5 \sigma$ ``effect.''  In a subsequent measurement of the top quark cross section~\cite{topCrossSection}, three candidate events were seen with an expected background of $0.21 \pm 0.16$, an excess corresponding to a $2.75 \sigma$ ``effect.''  Using \Sherlock, we find $\scriptP= 0.03$ in the $e\mu\met jj$ sample, a $1.9 \sigma$ ``effect,'' when complete ignorance of the top quark is feigned.  When we take into account the fact that we have also searched in all of the final states listed in Table~\ref{expectedBackgrounds1}, we find $\gothicP=0.11$, a $1.2 \sigma$ ``effect.''  The difference between the $2.75\sigma$ ``effect'' seen with a dedicated top quark search and the $1.2\sigma$ ``effect'' that \Sherlock\ reports in $e\mu X$ lies partially in the fact that \Sherlock\ is not optimized for $t\bar{t}$; and partially in the careful accounting of the many new physics signatures that \Sherlock\ considered in addition to $t\bar{t}$ production, and the correspondingly many new physics signals that \Sherlock\ might have discovered.  

\subsection{Search for physics beyond the standard model}
\def \twiddleScriptPValue {0.72}

In this section we present \Sherlock's results for the case in which all standard model and instrumental backgrounds are considered in the background estimate: $Z/\gamma^*\rightarrow\tau\tau$, fakes, $WW$, and $t\bar{t}$.  The results are shown in Table~\ref{tbl:tbl03} and Fig.~\ref{fig:data_mapping_nosig}.  We observe excellent agreement with the standard model.  We conclude that these data contain no evidence of new physics at high $p_T$, and calculate that a fraction $\gothicP= \twiddleScriptPValue$ of hypothetical similar experimental runs would produce a more sig\-nif\-i\-cant excess than any observed in these data.  Recall that we are sensitive to new high $p_T$ physics with production cross section $\times$ branching ratio into this final state as described in Sec.~\ref{section:RoughSensitivityScale}.

\begin{table}[htb]
\centering
\begin{tabular}{cc}
Data set  	& $\scriptP$ \\ \hline
$e\mu\met$   	& 0.14	\\
$e\mu\met j$  	& 0.45	\\
$e\mu\met jj$ 	& 0.31	\\
$e\mu\met jjj$  & 0.71	\\ \hline
\raisebox{-.6ex}{$\gothicP$}	& \twiddleScriptPValue  \\ 
\end{tabular}
\caption{Summary of results on all final states within $e\mu X$ when all standard model backgrounds are included.  The unpopulated final states (listed in Table~\ref{expectedBackgrounds2}) have $\scriptP=1.0$; these final states are included in the calculation of $\gothicP$.  We observe no evidence for the presence of new high $p_T$ physics.}
\label{tbl:tbl03}
\end{table}

{\dofig {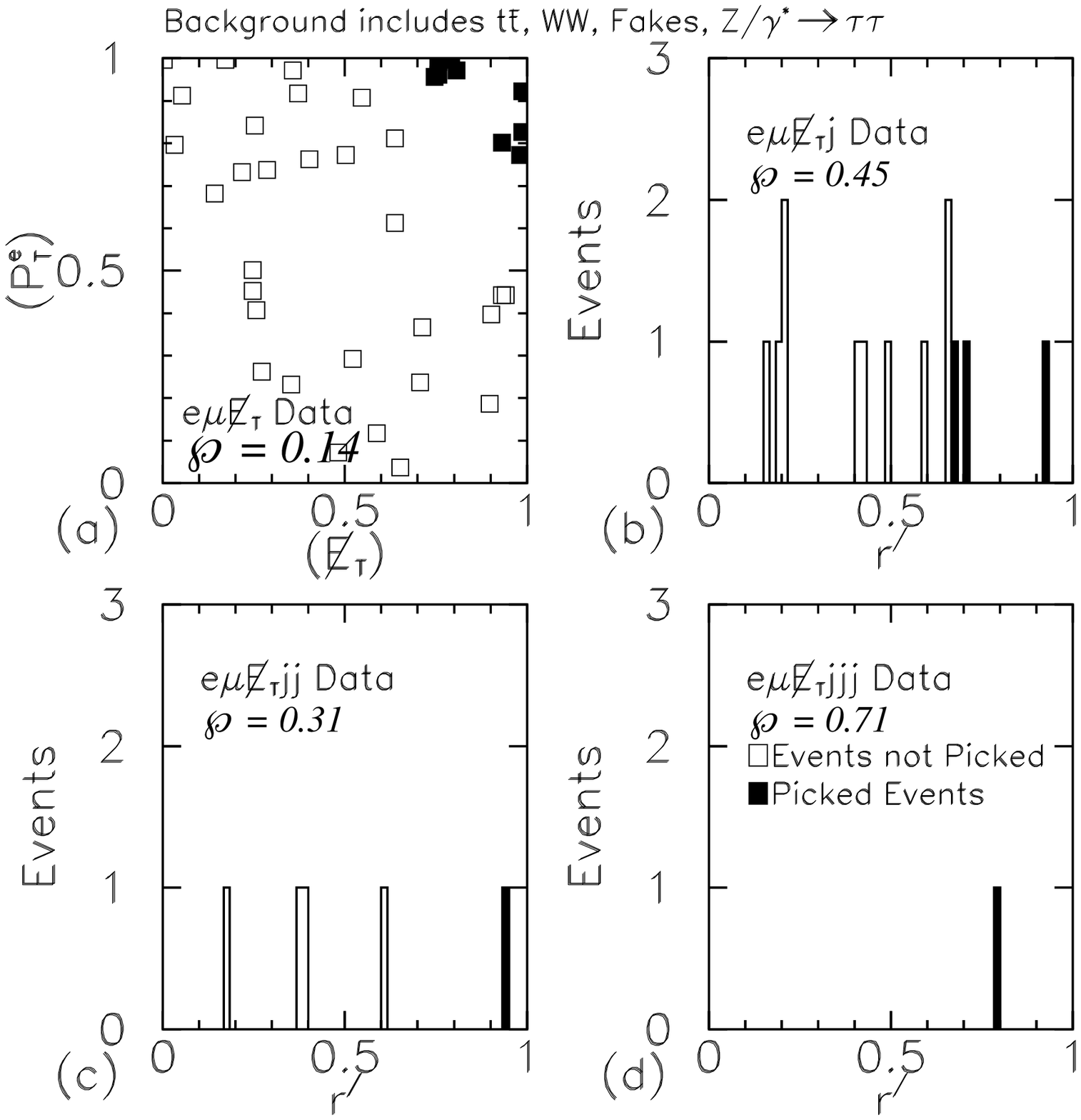} {3.5in} {Positions of the data points following the transformation of the background from $Z/\gamma^*\rightarrow\tau\tau$, fakes, $WW$, and $t\bar{t}$ sources in the space of variables in Table~\ref{tbl:VariableRules} to a uniform distribution in the unit box.  The darkened points define the region that \Sherlock\ chose.  We find $\gothicP=\twiddleScriptPValue$, and distributions that are all roughly uniform and consistent with background.  No evidence for new high $p_T$ physics is observed.} {fig:data_mapping_nosig}}


\section{Conclusions}

We have developed a quasi-model-independent technique for searching for the physics responsible for stabilizing electroweak symmetry breaking.  Our prescription involves the definition of final states and the construction of a rule that identifies a set of relevant variables for any particular final state.  An algorithm (\Sherlock) systematically searches for regions of excess in those variables, and quantifies the significance of any observed excess.  This technique is sufficiently {\em a priori} that it allows an {\em ex post facto}, quantitative measure of the degree to which curious events are interesting.  After demonstrating the sensitivity of the method, we have applied it to the set of events in the semi-inclusive channel $e\mu X$.  Removing $WW$ and $t\bar{t}$ from the calculated background, we find indications of these signals in the data.  Including these background channels, we find that these data contain no evidence of new physics at high $p_T$.  A fraction $\gothicP= \twiddleScriptPValue$ of hypothetical similar experimental runs would produce a more significant excess than any observed in these data.

\section*{Acknowledgments}

%
We thank the staffs at Fermilab and at collaborating institutions 
for contributions to this work, and acknowledge support from the 
Department of Energy and National Science Foundation (USA),  
Commissariat  \` a L'Energie Atomique and
CNRS/Institut National de Physique Nucl\'eaire et 
de Physique des Particules (France), 
Ministry for Science and Technology and Ministry for Atomic 
   Energy (Russia),
CAPES and CNPq (Brazil),
Departments of Atomic Energy and Science and Education (India),
Colciencias (Colombia),
CONACyT (Mexico),
Ministry of Education and KOSEF (Korea),
CONICET and UBACyT (Argentina),
A.P. Sloan Foundation,
and the Humboldt Foundation.
%

\appendix

\section{Further comments on variables}
\label{section:MoreOnVariables}

We have excluded a number of ``standard'' variables from the list in Table~\ref{tbl:VariableRules} for various reasons:  some are helpful for specific models but not helpful in general; some are partially redundant with variables already on the list; some we have omitted because we felt they were less well-motivated than the variables on the list, and we wish to keep the list of variables short.  Two of the perhaps most significant omissions are invariant masses and topological variables.

\begin{itemize}
\item{Invariant masses:  If a particle of mass $m$ is produced and its decay products are known, then the invariant mass of those decay products is an obvious variable to consider.  $M^T_{\ell \nu}$ and $M_{\ell^+\ell^-}$ are used in this spirit to identify $W$ and $Z$ bosons, respectively, as described in Sec.~\ref{section:SearchStrategy}.  Unfortunately, a non-standard-model particle's decay products are generally not known, both because the particle itself is not known and because of final state combinatorics, and resolution effects can wash out a mass peak unless one knows where to look.  Invariant masses turn out to be remarkably ineffective for the type of general search we wish to perform.  For example, a natural invariant mass to consider in $e\mu\met jj$ is the invariant mass of the two jets ($m_{jj}$); since top quark events do not cluster in this variable, they would not be discovered by its use.  A search for any {\em particular} new particle with known decay products is best done with a dedicated analysis.  For these reasons the list of variables in Table~\ref{tbl:VariableRules} does not include invariant masses.}
\item{Shape variables: Thrust, sphericity, aplanarity, centrality, and other topological variables often prove to be good choices for model-specific searches, but new physics could appear in a variety of topologies.  Many of the processes that could show up in these variables already populate the tails of the variables in Table~\ref{tbl:VariableRules}.  If a shape variable is included, the choice of that particular variable must be justified.  We choose not to use topological variables, but we do require physics objects to be central (e.g., $\abs{\eta_j}<2.5$), to similar effect.}
\end{itemize}

\section{Transformation of variables}
\label{section:VariableTransformationAppendix}

The details of the variable transformation are most easily understood in one dimension, and for this we can consider again Fig.~\ref{fig:pointOnTail}.  It is easy to show that if the background distribution is described by the curve $b(x)=\frac{1}{5}e^{-x/5}$ and we let $y=1-e^{-x/5}$, then $y$ is distributed uniformly between 0 and 1.  The situation is more complicated when the background is given to us as a set of Monte Carlo points that cannot be described by a simple parameterization, and it is further complicated when these points live in several dimensions.

There is a unique solution to this problem in one dimension, but an infinity of solutions in two or more dimensions.  Not all of these solutions are equally reasonable, however --- there are two additional properties that the solution should have.

\begin{itemize}
\item{Axes should map to axes.  If the data live in a three-dimensional space in the octant with all coordinates positive, for example, then it is natural to map the coordinate axes to the axes of the box.}
\item{Points that are near each other should map to points that are near each other, subject to the constraint that the resulting background probability distribution be flat within the unit box.}
\end{itemize}

This somewhat abstract and not entirely well-posed problem is helped by considering an analogous physical problem:

\begin{quote}
The height of the sand in a $d$-dimensional unit sandbox is 
given by the function $b(\vec{x})$, where $\vec{x}$ is a 
$d$-component vector.  (The counting of dimensions is such 
that a physical sandbox has $d = 2$.)  We take the 
$d$-dimensional lid of the sandbox and squash the sand flat.  
The result of this squashing is that a sand grain at position 
$\vec{x}$ has moved to a new position $\vec{y}$, and the new 
function $b'(\vec{y})$ describing the height of the sand is a 
constant.  Given the function $b(\vec{x})$, determine the 
mapping $\vec{x} \rightarrow \vec{y}$.
\end{quote}

For this analogy to help, the background first needs to be put ``in the sandbox.''  Each of the background events must also have the same weight (the reason for this will become clear shortly).  The background probability density is therefore estimated in the original variables using Probability Density Estimation~\cite{PDE}, and $M$ events are sampled from this distribution.  

These $M$ events are then put ``into the sandbox'' by transforming each variable (individually) into the interval $\left[0,1\right]$.  The new variable is given by
\begineq
x_j \rightarrow x'_j = \frac{1}{M} \int_{-\infty}^{x_j}{\sum_{i=1}^{M}{ \frac{1}{\sqrt{2\pi}\sigma_j h} \exp\left(-\frac{(t - {\mu_i}_j)^2}{2\sigma_j^2 h^2}\right)} \, dt},
\endeq
where ${\mu_i}_j$ is the value of the $j^{th}$ variable for the $i^{th}$ background event, $\sigma_j$ is the standard deviation of the distribution in the $j^{th}$ variable, and $h = M^{-\frac{1}{d+4}}$, where $d$ is the dimensionality of the space.

The next step is to take these $M$ events and map each of them to a point on a uniform grid within the box.  The previous paragraph defines a mapping from the original variables into the unit sandbox; this step defines a mapping from a lumpy distribution in the sandbox to a flat distribution.  The mapping is continued to the entire \mbox{space} by interpolating between the sampled background events.

The mapping to the grid is done by first assigning each sampled background point to an arbitrary grid point.  Each background point $i$ is some distance $d_{ij}$ away from the grid point $j$ with which it is paired.  We then loop over pairs of background points $i$ and $i'$, which are associated with grid points $j$ and $j'$, and swap the associations (associate $i$ with $j'$ and $i'$ with $j$) if $\max(d_{ij}, d_{i'j'}) > \max(d_{i'j}, d_{ij'})$.  This looping and swapping is continued until an equilibrium state is reached.

\section{Region criteria}
\label{section:RegionCriteriaAppendix}

In Sec.~\ref{section:RegionCriteria} we introduced the formal notion of {\em region criteria} --- properties that we require a region to have for it to be considered by \Sherlock.  The two criteria that we have decided to impose in the analysis of the $e\mu X$ data are {\em Isolation} and {\em AntiCornerSphere}.

\paragraph{Isolation}

We want the region to include events that are very close to it.  We define $\xi=\frac{1}{4}N_{\rm{data}}^{-\frac{1}{d}}$ as a measure of the mean distance between data points in their transformed coordinates, and call a region {\it isolated} if there exist no data points outside the region that are closer than $\xi$ to a data point inside the region.  We generalize this boolean criterion to the interval $[0,1]$ by defining
\begineq
c_R^{\rm {\it Isolation}} = \min{\left(1,\frac{\min{\abs{ (\vec{x})^{\rm in} - (\vec{x})^{\rm out} }}}{2\xi}\right)},
\endeq
where the minimum is taken over all pairwise combinations of data points with $(\vec{x})^{\rm in}$ inside $R$ and $(\vec{x})^{\rm out}$ outside $R$.

\paragraph{AntiCornerSphere}

One must be able to draw a sphere centered on the origin of the unit box containing all data events outside the region and no data events inside the region.  This is useful if the signal is expected to lie in the upper right-hand corner of the unit box.  We generalize this boolean criterion to the interval $[0,1]$ as described in Sec.~\ref{section:RegionCriteria}.

\vskip 12pt
\noindent A number of other potentially useful region criteria may be imagined.  Among those that we have considered are {\em Connectivity}, {\em Convexity}, {\em Peg}, and {\em Hyperplanes}.  Although we present only the boolean forms of these criteria here, they may be generalized to the interval $[0,1]$ by introducing the scale $\xi$ in the same spirit as above.

\paragraph{Connectivity}

We generally expect a discovery region to be one connected subspace in the variables we use, rather than several disconnected subspaces.  Although one can posit cases in which the signal region is not connected (perhaps signal appears in the two regions $\eta>2$ and $\eta<-2$), one should be able to easily avoid this with an appropriate choice of variables.  (In this example, we should use $\abs{\eta}$ rather than $\eta$.)  We defined the concept of neighboring data points in the discussion of regions in Sec.~\ref{section:Voronoi}.  A {\it connected region} is defined to be a region in which given any two points $a$ and $b$ within the region, there exists a list of points ${ p_1 = a, p_2, \ldots, p_{n-1}, p_n = b}$ such that all the $p_i$ are in the region and $p_{i+1}$ is a neighbor of $p_i$.  

\paragraph{Convexity}

We define a {\em non-convex} region as a region defined by a set of $N$ data points $P$, such that there exists a data point $\hat{\vec{p}}$ not within $P$ satisfying
\begin{eqnarray}
\sum_{i=1}^{N}{\vec{p}_i\lambda_i} = \hat{\vec{p}}\\
\sum_{i}{\lambda_i}=1 \\
\lambda_i \ge 0 \; \; \; \; \forall \; i,
\end{eqnarray}
 for suitably chosen $\lambda_i$, where $\vec{p}_i$ are the points within $P$.  A convex region is then any region that is not non-convex; intuitively, a convex region is one that is ``roundish,'' without protrusions or intrusions.

\paragraph{Peg}

We may want to consider only regions that live on the high tails of a distribution.  More generally, we may want to only consider regions that contain one or more of $n$ specific points in variable space.  Call this set of points $\tilde{x}_i$, where $i={1,\ldots,n}$.  We transform these points exactly as we transformed the data in Sec.~\ref{section:Regions} to obtain a set of points $\tilde{y}_i$ that live in the unit box.  A region $R$ is said to be {\it pegged} to these points if there exists at least one $i\in{1,\ldots,n}$ such that the closest data point to $\tilde{y}_i$ lies within $R$.

\paragraph{Hyperplanes}

Connectivity and Convexity are criteria that require the region to be ``reasonably-shaped,'' while Peg is designed to ensure that the region is ``in a believable location.''  It is possible, and may at times be desirable, to impose a criterion that judges both shape and location simultaneously.  A region $R$ in a $d$-dimensional unit box is said to satisfy {\em Hyperplanes} if, for each data point $p$ inside $R$, one can draw a $(d-1)$-dimensional hyperplane through $p$ such that all data points on the side of the hyperplane containing the point $\vec{1}$ (the ``upper right-hand corner of the unit box'') are inside $R$.

\vskip 12pt
\noindent More complicated region criteria may be built from combinations and variations of these and other basic elements.

\section{Search heuristic details}
\label{section:searchHeuristicDetails}

The heuristic \Sherlock\ uses to search for the region of greatest excess may usefully be visualized as a set of rules for an amoeba to move within the unit box.  We monitor the amoeba's progress by maintaining a list of the most interesting region of size $N$ (one for each $N$) that the amoeba has visited so far.  At each state, the amoeba is the region under consideration, and the rules tell us what region to consider next.  

The initial location and size of the amoeba is determined by the following rules for {\it seeding}:
\begin{enumerate}
\item{If we have not yet searched this data set at all, the starting amoeba fills the entire box.}
\item{Otherwise, the amoeba starts out as the region around a single random point that has not yet inhabited a ``small'' region that we have considered so far.  We consider a region $R$ to be small if adding or removing an individual point can have a sizeable effect on the $p_N^R$; in practice, a region is small if $N$ \raisebox{-0.6ex}{$\stackrel{<}{\sim}$} $20$.}
\item{If there is no point that has not yet inhabited a small region that we have considered so far, the search is complete.}
\end{enumerate}

At each stage, the amoeba either {\it grows} or {\it shrinks}.  It begins by attempting to grow.  The rules for growth are:

\begin{enumerate}
\item{Allow the amoeba to encompass a neighboring data point.  Force it to encompass any other data points necessary to make the expanded amoeba satisfy all criteria.  Check to see whether the $p_N^R$ of the expanded amoeba is less than the $p_N^R$ of the region on the list of the same size.  If so, the amoeba has successfully grown, the list of the most interesting regions is updated, and the amoeba tries to grow again.  If not, the amoeba shrinks back to its former size and repeats the same process using a different neighboring data point.}
\item{If the amoeba has tried all neighboring data points and has not successfully grown, it shrinks.}
\end{enumerate}
The rules for shrinking are:

\begin{enumerate}
\item{Force the amoeba to relinquish the data point that owns the most background, subject to the requirement that the resulting shrunken amoeba be consistent with the criteria.}
\item{If the amoeba has shrunk out of existence or can shrink no further, we kill this amoeba and reseed.}
\end{enumerate}  

The result of this process is a list of regions of length $N_{\rm data}$ (one region for each $N$), such that the $N^{th}$ region in the list is the most interesting region of size $N$ found in the data set.

\bibliographystyle{unsrt}

\end{document}